# *T*ransforming *A*utomatically *B*PMN Models to *S*mart *C*ontracts with Nested Trade Transactions (TABS+)

Transforming BPMN Models to Smart Contracts with Support of Nested Transactions

Christian, G., Liu

Faculty of Computer Science, Dalhousie University, Chris.Liu@dal.ca

Peter Bodorik

Faculty of Computer Science, Dalhousie University, Peter.Bodorik@dal.ca

Dawn Jutla

Sobey School of Business, Saint Mary's University, Dawn.Jutla@gmail.com

Development of blockchain smart contracts is more difficult than mainstream software development because the underlying blockchain infrastructure poses additional complexity. To ease the developer's task of writing smart contract, we also use Business Process Model and Notation (BPMN) modeling to describe application requirements for trade of goods and services and then transform automatically the BPMN model into the methods of a smart contract. In our previous research we described our approach and a tool to *Transform Automatically BPMN models into Smart contracts* (**TABS**). In this paper, we describe how the TABS approach is augmented with the support for a BPMN trade transaction by several actors. Our approach analyzes the BPMN model to determine which patterns in the BPMN model are suitable for use as trade transactions and show those patterns to the developer who decides which ones should be deployed as trade transactions. We describe how our approach automatically transform the BPMN model into smart contract that provides a transaction mechanism to enforce the transactional properties of the nested transactions. Our approach greatly reduces the developer's task as synchronization of collaborative activities is provided by our approach, so that the developer needs to code only isolated tasks with well-defined inputs and outputs. We also overview the TABS+ tool we built as a proof of concept to show that our approach is feasible, and we provide estimates on the cost of supporting the nested BPMN trade transactions.

**ACM Reference Format:**

## 1 INTRODUCTION

Nakamoto's introduction to the Bitcoin cryptocurrency in 2008 [42], and its subsequent rise (Marr 2017 [37]), spurred great interest in cryptocurrencies and blockchains. Further blockchain platforms, such as Hyperledger fabric and Ethereum, followed the Bitcoin blockchain, and many of them succeeded while many have failed. The concept of a smart contract (Szabo 1996 [49]; Buterin 2015 [10]), which is stored on the blockchain ledger, emerged as a collection of methods written in a Turing-complete high-level language. As the blockchain infrastructure utilizes cryptographic concepts and methods to provide the blockchain's desirable properties of trust, immutability, availability, and transparency, amongst others, the smart contact also benefits from these properties as it is stored on the blockchain itself when it is deployed.

However, as a new technology, blockchain and their smart contracts pose new challenges in both the blockchain infrastructure and in developing smart contracts and applications that use them. Thus, in addition to tackling the blockchain infrastructure issues of scalability, transaction throughput, and high costs, the development of smart contracts also received much attention by the research and development communities as can be seen by many literature surveys on the topic, such as in Taylor 2019 [50], Khan 2021 [26], Vacca 2021 [54], Belchior 2021 [1], Saito 2016 [46], Garcia-Garcia 2020 [19], Lauster 2020 [27], and Levasseur 2021 [28].

Khan 2021 [26] classifies research on blockchain into the categories of improvement and usage, in which the improvement category includes research and tools to improve smart contracts either when writing them or once they exist. Improvement approaches for writing smart contracts generally rely on defining a programming language that facilitates creation of smart contract with desirable properties. However, although these approaches lead to smart contract programs with formally proven safety properties, they have not been adopted in practice as they exhibit, or are perceived to exhibit by software developers, complexity that deters their adoption by developers who ultimately decide how well-accepted a programming language is. To alleviate such issues, some researchers use the well-known and used Business Process Management Notation (BPMN) to model the application requirements and transform a BPMN model automatically into the methods of a smart contract(s) (Weber 2016 [55], López-Pintado 2019 [34-35], Tran 2018 [51], Bodorik 2023 [4], Liu 2021(a) [29]).

### 1.1 Motivation

Trade of goods and services, as opposed to securities, includes collaborative activities that arise with a sale, trade and finance of goods and/or services between two or more parties. Trade activities can be classified into single-action transactions, such as payment on an account, transfer of funds, or purchase of an item. Such transactions are by far most frequent and blockchains that support such transactions concentrate on security and throughput of transactions. However, in trade and finance of goods and services, multi-step transactions that involve a number of participants are a norm. For instance, a sale of a product may involve price negotiation and arrangements for delivery, insurance, and payments. Thus, a sale of a product may involve seller and a buyer negotiating the price, a third party to hold escrow account for payment, transporters/shippers  for the delivery of the purchased goods, insurance for risk mitigation, customs for clearance of goods, regulators when crossing international or state/provincial/municipal boundaries or when goods contain dangerous substances, and banks may be involved to provide financial instruments supporting the purchase, instruments such as a letter of credit that guarantees the payment of goods to the seller's bank upon receiving the goods or services. We shall refer to such transactions as trade transactions, and we note that they may be long term, e.g., spanning, minutes, hours, or days. Performing some activities may depend on other activities, such as shipment of the product should occur only after the byer deposits funds for payment in an escrow account, while some are independent of some other activities within the same trade transactions. Thus, in the context of trade and finance, the concept of a transaction may be loosely defined as a collection of activities that need to be performed with certain constraints amongst various activities, while some activities may have no constraints in relation of some other activities within a trade transaction.

In addition to the constraints arising from the blockchain infrastructure, such as limited stack space, blockchain smart contracts impose another restriction on writing smart contracts: A blockchain transaction is defined as updates made by an execution of any one of the methods of the smart contract. Instead of the developer creating a transaction, which contains sub-transactions and then developing sub-transaction, the developer must first determine the individual base blockchain transactions, i.e., individual smart contract methods that write the blockchain transactions, and then turn her/his attention to writing methods that synchronize the collaborative activities.

To ease the development of the smart contract, as in other research [Weber 2016 [55], López-Pintado 2019 [34-35], Tran 2018 [51], , Mendling 2018 [40], Loukil 2021 [36]), we developed a methodology that starts with using a well-known BPMN to model the application requirements and then transforms the BPMN model automatically into the methods of a smart contract(s). In [Bodorik 2023 [4], Liu 2021 [29-31], Liu 2022 [32]), we describe our approach, and the **TABS** tool (**T**ool to **A**utomatically Transform a **B**PMN Model to **S**mart Contract Methods) as a **P**roof **o**f **C**oncept (**PoC**), to show the feasibility of our approach, however, without the support of trade transactions. Our TABS approach is very different form the other approaches to transforming BPMN models into the methods of a smart contract as we use multi-modal *Discrete Event (DE) – Hierarchical State Machine (HSM)* modeling (*DE-HSM* modeling). The significance is that the DE-HSM modeling enables a graph representation of the distributed blockchain application that, in turn, facilitate subsequent analysis that results in finding appropriate patterns that are isolated from each other and hence do not interfere with other concurrent activities. We exploit this property in the TABS approach to facilitate sidechain processing for cost reduction, wherein a part of the smart contract computation is performed on a sidechain. The cost reduction is achieved if the processing on a sidechain (e.g., Solana or Quorum) is sufficiently cheaper than processing on the mainchain (e.g., the expensive public Ethereum blockchain) to compensate for overhead cost to facilitate the sidechain processing. In this paper, we describe how we extend the TABS approach and its tool into the TABS+ approach and tool that provide support for trade transactions that may be nested. We describe how to (i) identify BPMN patterns that are suitable for defining as transactions and (ii) how to automatically create a transaction mechanism to enforce the properties of long-term transactions that may be nested. The main objective is to ease the developer's effort in creating smart contracts.

## 1.2 Goal and Objectives

The main goal of this paper is to describe how the TABS approach is augmented to support defining and supporting trade transactions, which may be nested, to create a **TABS+** approach and tool that provide a mechanism (a) to identify BPMN trade transaction in the BPMN model and (b) to automatically transform the BPMN model with its transactions into smart contract methods that include functionality to support the trade transactions that may be nested. The specific objectives include:

– Transform BPMN model into a DE-HSM model that is then analyze it to find patterns that are suitable as candidates for trade transactions while also supporting nesting of transactions.
– Augment the TABS approach to support automated transformation of a BPMN model, which includes nested trade transactions, by including in the smart contract methods a transaction mechanism to support the nested trade transactions.
– Develop a PoC to show that the proposed approach is feasible by incorporating the support of nested trade transactions into the TABS tool, thus creating TABS+ tool.
– Provide an estimate of overhead in supporting the nested BPMN trade transaction using the TABS+ approach.

However, it should be kept in mind that the main goal of the TABS, and hance the TABS+ approach, is to support automated transformation of BPMN models into smart contracts with minimal effort on the software developer responsible for the creation of the smart contract(s).

## 1.3 Contributions

It was already reported above that the TABS approach, which provides for an automated transformation of a BPMN model into the methods of a smart contract with the support for sidechain processing, was described in (Bodorik 2023 [4]) with earlier work in (Bodorik 2021 [3], Liu 2021 [29-31], Liu 2022 [32])]. Before we describe contributions, we need to stress that the overall approach we adopt is to simplify the developer's task in creating smart contracts. The process starts with creation of the BPMN model. The model may be prepared by a business analyst or the software developer (see Section 2 for justification). The software developer responsible for creation of the smart contract performs the following actions to transform a BPMN model into the methods of a smart contract:

– Create the BPMN model using the TABS+ tool or import the model into the TABS+ tool.
– Guide the tool to transform the BPMN model into a DE-FSM model, which consists of clicking on a button/tab to perform transformation. The system analyzes the DE-HSM model to find suitable transactions that may be nested. The results that are mapped back to the BPMN model and shown to the developer graphically in the form of BPMN patterns.
– Review the results and identify those patterns that should be deployed and supported as transactions of which results are certified by the participants, and which are recorded, together with the certifications, on the blockchain.
– Provide the code/scripts for isolated tasks.
– Review the system provided estimates of the costs due to the overhead of the automatically generated nested transactional mechanisms and then select the transaction mechanism to be deployed.
– The tool asks the modeler/developer on which blockchain(s) should the smart contracts be deployed and then creates the smart contracts and deploys them on the selected blockchain(s).

The developer thus does not need to decompose the application into individual transactions; such decomposition is performed by the process transforming the BPMN model to the methods of smart contract. As the developer does not need to worry about the synchronization of the collaborative activities, her/his role is reduced primarily to coding of individual and isolated tasks. Most of the developer's effort is in providing/writing code for individual tasks represented in the BPMN model, tasks that become parts of a smart contract. However, each task is an individual method that is isolated, in that it does not interact or interfere with other tasks or smart contract methods, and that has well-defined inputs and outputs. The developer is provided with information gathered from the BPMN diagram that informs the developer about the input parameters for each individual task represented by a smart contract method, the purpose of the method/task, as well as information about the output parameters that are the result of the task execution, and their subsequent use.

In this paper we describe how the TABS approach is amended to support transformation of BPMN models into smart contracts with the support for nested trade transactions, resulting in the TABS+ approach and tool. More specifically:

– We show that the transformation of a BPMN model into DE-HSM model and then DE-FMS model allows for the analysis of the model's graph representation to find patterns that are subgraphs in the DE-FSM model, referred to as *Single Entry Single Exit* (SESE) subgraphs, that have localization of execution properties that make them suitable candidates for BPMN trade transactions.

– We augment the TABS tool and its approach to support BPMN trade transactions that are transformed into smart contracts that contain transactions that *span multiple methods* and are *nested* - thus creating TABS+ approach and tool.

– We report on costs associated with creation of a transaction mechanism that supports the nested BPMN trade transactions.

In short, the design of smart contracts to support **D**istributed blockchain **app**lications (**Dapps**) in the vertical of trade and finance of goods and services is greatly simplified using our approach.

### 1.4   Outline

Section 2 overviews the background information for this paper. It briefly overviews **H**ierarchical **S**tate **M**achines (**HSM**s) and multi-modal modeling. Considering that the support of the nested trade transactions is based on the TABS approach and tool, the background section also overviews the TABS approach to automated transformation of a BPMN model into the methods of smart contracts with the support of sidechain processing.

Section 3 reviews the main features of transformations from BPMN to smart contract methods and shows (i) how the approach to transformations and the TABS tool are augmented to support defining BPMN trade activities/transactions for a BPMN model, and (ii) how to create a mechanism to support the trade transaction in addition to supporting the sidechain processing. Finally, the section addresses how a transaction mechanism is automatically generated to support nested transactions.

Section 4 presents the description of the TABS+ tool and reports on experimental evaluation to estimate the cost of supporting nested trade transaction. The final section provides limitations and related work, and future work and concluding remarks.

## 2  BACKGROUND

We first briefly overview BPMN modeling and describe a simple BPMN use case that had been used also by other research in reports dealing with BPMN model transformation into a smart contract. We then overview HSMs and DE-HSM multi-modal modeling and then follow it by a high-level overview of the transformations performed in the TABS approach.

### 2.1  BPMN – Business Process Management Notation

BPMN was developed by the OMG organization (Business Process Model and Notation (BPMN), Version 2.0. 2023 [6-9]) with the objective of BPMN models to be understandable by all business users, from business analysts, through technical developers implementing the processes, to people managing those processes. It is viewed as a de-facto standard for describing business processes. That it has been adopted in practice is demonstrated by the many software platforms available that provide for modeling of business applications with the objective to automatically create an executable application from the BPMN model. For instance, Oracle Corporation uses BPMN to describe an application and transform into a blueprint of processes expressed in an executable Business Process Execution Language (BPEL) (Dijkman 2008 [16]), wherein the blueprint represents the logic of the application in terms of concurrent processes and their interactions, while details of individual tasks are supplied by implementors. Another example is Camunda software platform that is also used to develop a BPMN model that is then transformed automatically into a Java application (Deehan 2021 [14]).

### 2.2  FSMs, Hierarchical State Machines (HSMs), and Multi-modal Modeling

Because FSM modeling has been used frequently in the design and implementation of software, the FSM modeling has been expanded with features, such as a guard along an FSM transition, to specify a Boolean condition on the state's variables that must evaluate to true for the transition to take place. In the late 80's, FSMs were extended with the concept of hierarchy to address the issues of reuse of patterns, leading to Hierarchical State machines (HSMs) that may contain states that are themselves other FSMs. Harel (1987) [22] showed that FSMs can be combined hierarchically: A single hierarchical state at one level can be in several states concurrently as represented by an FSM(s) in a lower level of the hierarchy, and FSMs may also be combined leading to concurrent FSMs. An HSM can be defined using induction as follows (due to (Harrel 1987 [22])] and described in (Girault 1999 [20], Yannakakis 2000 [56], and others):

In the base case, an FSM is a hierarchical machine. Suppose that M is a set of HSMs. If F is an FSM with a set of states, S, and there is a mapping function f: S->M, then the triple (F, M, f) is an HSM. Each state, s ∈ S, that represents an HSM is replaced by its mapping (f(s)). HSMs recognize the same language as its corresponding flattened FSM. HSMs do not increase expressiveness of FSMs, only succinctness in representing them.

Girault et al. (1999) [20] describe how HSM modeling can be combined with concurrency semantics of several models, including communicating sequential processes (Hoare 1987 [23]) and discrete events (Cassandras 1993 [12]). Girault et al. (1999) [20] describe how an HSM model can represent a module of a system under a concurrency model that is applicable only if the system is in that state. This enables representation of a subsystem using a particular concurrency model that may be nested within a hierarchical state of a higher-level FSM. This may be used in multi-modal modeling, in which different (hierarchical) states may be combined with different concurrency models that are best suitable for modelling of concurrent activities for that particular state. We exploit the concept of multi-modal modeling to allow the designer to model concurrent, but independent activities, by concurrent FSMs at the lower level of hierarchy. In (Liu 2022 [32]), we showed that a multi-modal DE-HSM model may be used to model trade applications. We also showed that it is possible to automatically transform a DE-HSM model of such an application into methods of a smart contract deployed on a blockchain.

A DE-HSM has external inputs, and it produces outputs. The model represents how external inputs form inputs to the sub-models and how those sub-models are interconnected to produce the final output. However, if the sub-model's interconnection is such that there are no loops, then the model can be viewed as a zero-delay model (Yannakakis 2000 [56]) in which the individual DE queues may be combined into one DE queue.

Feedback loops in a DE-HSM arise due to feedback loops in the BPMN model from which the DE-HSM model is derived. None of the approaches, to producing smart contract from a BPMN model that were discussed in the introductory section, addressed the issues of the feedback loops in a BPMN model. Feedback loops cause various difficulties, such as deterministic semantics, and will be discussed further in a later section.

DE-HSM model is further transformed into the DE-FSM model by a straight-forward elaboration of a sub-model with its components. When we use the term DE-FSM model, we refer to a model that does not contain any hierarchical states, i.e., a model in which each HSM sub-model was "flattened out" by being replaced with its components.

## 2.3 TABS Approach to Transforming BPMN Models to Smart Contracts

As in other research, we also start with a BPMN model, but unlike the other approaches that transform the BPMN model directly into the methods of a smart contract, we use several transformations that result in a system represented as an interconnection of subsystems, with each subsystem represented using multi-modal modeling.

### 2.3.1 DE-HSM and DE-FSM Modeling to Represent Concurrency and Functionality

In DE-HSM multi-modal modeling, the functionality is represented using an HSM, while concurrency is modeled using Discrete Event (DE) modeling. Furthermore, each HSM model may be an interconnection of further DE-HSM sub-models (Bodorik 2021 [3], Bodorik 2023 [4], Liu 2022 [32]). Eventually, by elaborating each HSM with its internal representation, the system may be "flattened out" into an interconnection of DE-FSM sub-models, in which concurrency is represented using DE modeling, while functionality is represented by FSMs.

The DE-FSM model represents the flow of execution throughout the model such that many concurrent streams of executions are possible. From the BPMN start element, the flow of execution is forked using an inclusive gateway into one of n flows, one for each of the n actors, while the functionality of each flow is expressed using an FSM derived for each actor. Of course, there are likely to be forking gateways in the BPMN model that may result in forking a flow of execution into multiple executions of concurrent flows. And similarly, there are likely to be joining gateways in the BPMN model that join several execution-flows into one. Forking of execution flows is supported by using the concurrent FSMs.

Consider a forking inclusive gateway with guards A, B, C, and D as shown in Fig. 1(a). Outgoing edges of a BPMN inclusive fork gate represent the flows of executions/computations of multiple concurrent streams. The flow of execution for such a gate within a sub-model is represented by a control FSM, shown in Fig. 1(b), that evaluates the guards for multiple concurrent streams and results in four flows of executions in which concurrency is expressed using a DE model, while functionality is represented using an FSM, as shown in Fig. 1(c).

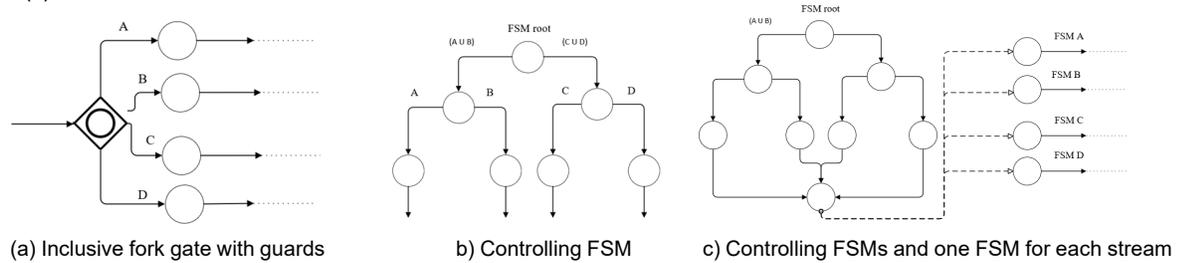

(a) Inclusive fork gate with guards        b) Controlling FSM        c) Controlling FSMs and one FSM for each stream

Fig. 1.   Inclusive fork gate resulting in a control FSM and one FSM for each concurrent process streams (adopted from (Liu 2021 [30])

### 2.3.2 Transforming a DE-FSM Model into the Methods of a Smart Contract

So how are the smart contract methods formed? Each actor has a smart contract method that acts as an interpreter and provides the execution environment for an actor as represented using the DE-FSM modeling. In other words, a smart contract method acts as an interpreter for executing the DE-FSM model that is stored within the monitor. Although the monitor script is the same for each actor, actor activities are represented by DE-FSM models that are unique for each actor. DE modeling uses a DE queue, a priority queue of events timestamped using the global clock supporting the Dapp and the smart contract. As the Dapp executes, it takes input from the actors, i.e., from the users or their systems acting on their behalf. The user input is examined to determine the event origin in terms of the BPMN diagram. This is then used to identify which of the actors is the event's target and the actor's smart contract method is invoked with the user input passed to the method as an input parameter.

The interpreter examines the origin of the input parameter in terms of the BPMN model, and it queues the event for execution into the DE queue of events for that sub-model. Once the event reaches the head of the queue, it is dequeued and "processed" by the interpreter method. The method examines the user input and determines to which of the concurrent FSMs for the actor's flow of execution it applies, and then it forwards the user input to that FSM. The FSM examines its current state and the input and fires, while producing a new state and generating its output. The output and the new state are examined and then the method may queue another new event into the DE queue or into a DE queue of another actor, or it may simply return - and then the process repeats until the queue is empty.

It should be noted that as there are no feedback loops within the DE-HSM and DE-FSM sub-models, all sub-models may share a single queue of events in DE modeling. Any input generated by the Dapp is first augmented with the information on the BPMN context (which BPMN element caused the method invocation) and is queued into the DE queue of events. The queued event eventually reaches the head of the DE queue and is dequeued and processed by the interpreter by invoking an FSM that results in another event being queued, or a task being executed or both, and the processes repeats with another DE queue element dequeued and processed. For instance, if the input for an FSM, which controls outgoing streams of an inclusive fork gate, is such as the one shown in Fig. 1(b), once the controlling FSM fires, its output is used to generate an event for each of the guards A, B, C, or D that was evaluated to true. However, the output from an FSM firing may also indicate that a specific task, corresponding to the BPMN task element, needs to be executed.

Fig. 2 shows the system architecture overview for the design phase and the execution phase. Monitor, shown as a part of the on-chain component in Fig.2, performs the role of interpreter. Referring to Fig. 2, in the <u>design phase</u>:

- The BPMN model is transformed into a DAG representation and then into an interconnection of DE-HSM sub-models.
- It is then transformed into an interconnection of DE-FSM sub-models and then into the methods of the smart contract.
- The smart contract with its methods is deployed and the API interface is installed that is invoked by the Dapp. The API interface examines the Dapp's input parameters and invokes an appropriate smart contract method with an input parameter containing the information about the Dapp's API call.

<u>Execution phase</u> in Fig. 2 includes two sub-phases: (1) Initialization and (2) Execution. Initialization deals with issues, such as authentication of the actors. Once the smart contract is deployed, execution is managed by the Monitor module. As long as the DE queue is not empty, the monitor repeatedly dequeues an element and examines it to determine which of the smart contract methods it belongs. It then invokes that smart contract method that determines which of its FSMs should be used to process the element, and it then invokes that FSM to fire with input being derived from the dequeued element. FSM fires, which results in a new FSM state and output from the FSM firing. The FSM's output and new state are examined in order to determine if a new DE element needs to be queued or if a task needs to be executed as a part of the FSM firing cycle.

The above description is a high-level overview of the TABS system architecture. It should be noted that many details have been omitted, including details how sidechain processing is supported by the system. For further details, see (Bodorik 2023 [4]).

## 3   TABS+:   TRANSFORMING A BPMN MODEL TO SMART CONTRACT METHODS WITH THE SUPPORT FOR BPMN TRADE TRANSACTIONS

In the first subsection, we review the major aspects of transforming a BPMN model into the methods of a smart contract as described in (Bodorik 2023 [4]). In the second subsection, we highlight how the graph representation of the multi-modal model is analyzed to find SESE subgraphs that are then transformed to the BPMN model representation that is shown to the developer who decides which of the SESE subgraphs should be transformed to transactions that satisfy the transactional properties. The SESE subgraphs selected by the developer as transactions are transformed into a separate smart contract method - thus satisfying the *isolation* property, referred to also as independence property in (Bodorik (2023) [4], (Liu 2023 [32]). Consequently, transactional mechanisms described in (Liu 2023 [33-34]) for mm-transaction support are applied on the methods generated from the SESE subgraphs selected by the developer.

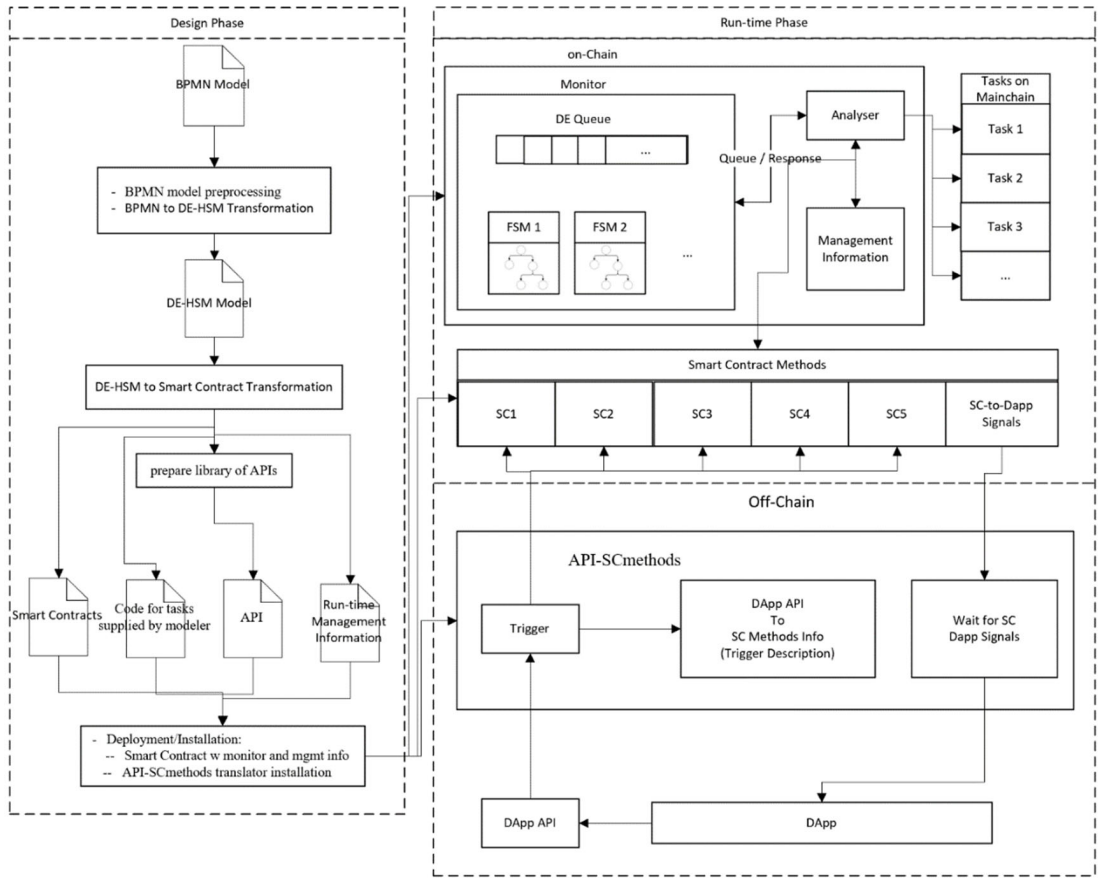

Fig. 2.   Architecture Overview (adopted from (Bodorik 2023 [4]))

The following subsections describe the overall process of transforming a BPMN model into the methods of a smart contract under the developer's guidance with the support for specifying nested BPMN trade transactions together with the description how the transactional support for the BPMN trade transactions, even when nested to support sub-transactions, is achieved.

### 3.1  TABS Transformations of a BPMN Model into Smart Contract Methods

We outline the TABS transformations, of a BPMN model into the methods of a smart contract, as presented in (Bodorik 2023 [4]) while omitting many details. The following subsections describe:

–   3.1.1 … Assumptions made on the BPMN model.
–   3.1.2 … Pre-processing and forming DAG representation of the BPMN model.
–   3.1.3 … How a DAG model is transformed into the multi-modal DE-HSM and then DE-FSM models.
–   3.1.4 … Transformation of the DE-FSM model into the methods of a smart contract(s).

### 3.1.1 Assumptions

We briefly highlight the assumptions made in (Bodorik 2023 [4]) on a BPMN model and its transformation into a smart contract:

_Assumptions on BPMN looping and parallel construct_: As in other work on transforming the BPMN model into a smart contract, we do not handle semantics of BPMN looping and parallel constructs so that the models are deterministic, and we handle these constructs as special cases in the final phase of the generation of the methods of smart contracts.

_Assumptions on BPMN converging gateways_: We make a simplification for an inclusive converging gateway in that we simply pass the token - we check the other pathways neither for enablement nor for a token arrival.

_No mm-transaction_: Multi-method transactions are not supported in the TABS tool (Bodorik 2023 [4]).

_Coverage of BPMN_: Not all BPMN symbols are supported. The list of currently supported symbols can be found in (Bodorik 2023 [4]).

_Task element_: The BPMN task element represents a self-contained task. The transformations prepare the method skeleton for each task element, and a task is invoked according to the BPMN model description. However, the code/script for the task must be supplied by the developer. This approach is standard in creating applications from a BPMN model. For instance, Oracle Corporation uses the same approach in its Oracle Business Process Analysis Suite when transforming a BPMN diagram into a blueprint for executable Business Process Execution Language (BPEL) processes (Dijkman 2008 [16]). A similar statement also applies to the Camunda platform that also asks the modeler to provide the script for each of the PBMN model's tasks (Camunda [11]).

_Off-chain storage_: We adopt the standard practice to store large objects/data off-chain and store on the mainchain only the hash-code of the data stored off-chain. Any time the object is retrieved by a smart contract, to ensure the immutability property of an object stored off-chain, its hash-code is recalculated and checked against the hash code stored on the blockchain. We facilitate off-chain storage by using IPFS, which is a distributed system for storing and accessing files, websites, applications, and data (Steichen 2018 [48]). IPFS storage is content addressable using the hash-code of the object. Storage in IPFS may be replicated and by controlling replication desirable resilience to storage failures may be achieved [41, 44].

### 3.1.2 BPMN Pre-processing and **D**irected **A**cyclic **G**raph (**DAG**) Generation

Once a BPMN model is created and expressed using its standard XML representation (BPMN 2.0 Introduction - Flowable Open Source Documentation (2022) [6]), the model is transformed into an equivalent BPMN model that is well-formed, wherein the well-formed BPMN model is defined as in (Dijkman 2008 [16]): (i) there is only one _start_ event; (ii) there is only one _end_ event; (iv) _fork/split decision gateways_ have one incoming flow and more than one outgoing flow (v) _join/merge gateways_ have one outgoing flow and more than one incoming flow; and (vi) there are _no data-based splits/forks or joins/merges_ – they are expressed using equivalent gateways constructs. The above conditions are not restrictive as any BPMN model can be transformed into an equivalent well-formed BPMN model as described in (Dijkman 2008 [16]). Furthermore, the execution flow from the _start event_ BPMN element is immediately forked, by an inclusive fork gateway, into n concurrent/parallel flows of executions, one for each of the actors. Similarly, the parallel streams for the actors are eventually joined, by a join gateway to end at vertex corresponding to the _end event_ BPMN element.

Following the preprocessing, the BPMN model is transformed into its DAG representation in which the edges represent the flows of execution. Using BPMN association data, the developer annotates the data flow along the edges with the data descriptions. Vertices represent either self-contained tasks, for which code is supplied by the developer and which consume the data along the incoming edges and output data on the outgoing edges, or elements that represent the emitters or receptors of information/events, such as gateways controlling the flow of execution or event generation/consumption, which also control or alter the flow of execution. Flows of executions may be concurrent as controlled by BPMN elements, such as gateways or event generators/receptors. As the BPMN model does not have looping under the stated assumptions, the graph representation of the BPMN model does not have any cycles either – hence a DAG representation.

### 3.1.3 DAG Transformation into DE-HSM Sub-models and then to DE-FSM Sub-models

We analyze the DAG representation of the BPMN model in order to find all SESE subgraphs. We show that due to the underlying graph being a DAG, the SESE subgraphs are also acyclic. As a SESE subgraph is acyclic, the subgraph may be analyzed using multi-modal modeling in which concurrency is modeled using DE modeling and functionality is expressed using HSMs. Furthermore, we show in (Bodorik 2023 [4]) that the DAG representation of the BPMN model can be represented as a collection of mutually exclusive SESE subgraphs, such that all DAG nodes appear exactly in one of the subgraphs and each edge either appears in one of the subgraphs or it connects two of the subgraphs. The SESE subgraphs have further properties that we discuss later when describing how the transactional mechanism is implemented to support BPMN trade transactions.

Each of the DE-HSM sub-models is "flattened out" by representing each HSM by its representation as an interconnection of its DE-HSM sub-models until each off the sub-models is a DE-FSM model. It is the DE-FSM model representation that is used to produce the methods of a smart contract(s).

### 3.1.4 Transforming DE-FSM Model into the Methods of a Smart Contract(s)

Each actor has a smart contract method that acts as a monitor/interpreter and controls the flow of execution for that actor according to the DE-FMS model for that actor. The flow of execution is represented by a DE queue of events that are continuously dequeued during the execution phase by the interpreter/monitor and processed by taking the dequeued element's information to determine which FSM should be fired with input derived from the dequeued DE element as described in the subsection 2.3. Once the smart contract(s) is generated, compiled and deployed, initialization is performed, and the smart contract is ready for invocation of its methods.

## 3.2   BPMN Trade Transactions: Properties and Specifying

Transformations described above are for the case when there are no BPMN trade transactions, and no sidechain processing is used. Before we discuss how to automatically create the transaction mechanism, the issue of which transactional properties need to be supported is addressed in the subsection 3.2.1. The subsection 3.2.2 describes a use case that is used for exposition purposes in this paper. The subsection 3.2.3 addresses how the developer identifies which BPMN patterns constitute trade transactions, while the subsection 3.2.4 shows how to create the transactional mechanism to support the trade transactions in an automated fashion.

*3.2.1 Transactional Properties for Trade Transactions*

To determine the transactional properties, we draw upon previous concepts of transactions in DBs and in blockchains.

<u>DB Transaction Properties</u>

A relational DB provides a mechanism that allows an application to issue the commands to begin and end a transaction. Any DB reads and write operations, also referred to as **C**reate, **R**ead, **U**pdate, **D**elete (**CRUD**) operations, between the two start and end transaction commands constitute a transaction. Furthermore, a DB transactional mechanism enforces the **A**tomicity, **C**onsistency, **I**solation, and **D**urability (**ACID**) properties for a transaction, referred to as DB transactional or ACID properties.

<u>Native Blockchain Transactions and Their Properties</u>

We use the term "native blockchain transaction" to refer to the usual definition of a blockchain transaction that is the result of access made by an execution of any of the individual smart contract methods that write to the ledger. Blockchains support the ACID properties for a native blockchain transaction: Ledger writes are not applied on the blockchain immediately but are collected into a block. Each transaction within a block, which is to be appended to the chain, is checked for consistency against existing transactions already recorded on the chain and against other transactions within the same block. Thus, in the DB-speak, the property of serializability, and hence *consistency,* of transactions is assured; or in the blockchain-speak, double spending is prevented. *Atomicity* and *isolation* are also assured as the transaction writes to the ledger are not visible to other transactions until the block is approved and appended to the chain, wherein the action of approving and appending a block to the chain is *atomic*. Finally, if there is no forking of the chain, *durability* is also supported as the chain is replicated.

<u>Specifying and Supporting Multi-method Blockchain Transactions and Their Properties</u>

Blockchains, however, do not provide any concept of a transaction that is the result of an execution of more than one method of a smart contract, whether it is executions of separate methods or repeated execution of the same method, or some combinations of the two. We refer to such a transaction as a ***m***ulti-***m***ethod (***mm***) transaction in (Liu 2023 [33]), where we described how to define an mm-transaction as a subset of the methods, of the smart contract, wherein the mm-transaction methods are *isolated/independent,* in that such methods do not interfere with each other. The isolation property prevents the transaction methods to make references to objects, or invoke methods, that are declared externally to the methods of the mm-transaction as that would violate the atomicity and isolation properties. If a transaction invokes an external method X, and the transaction is then aborted, the activities of the method X would also have to be aborted/negated or else the atomicity and isolation properties would be violated.

In addition to describing how a developer identifies an mm-transaction, (Liu (2023) [33]) also shows that the *ACID properties* must be supported, and that *access control* and *privacy* should be supported:

-   The *access control* property states that only the actors who are participants to the transactions may invoke the transaction methods.
-   The *privacy* property states that the transaction ledger updates should be visible only to the participants of the transaction while the transaction is in progress. It should be noted that in (Liu 2023 [33]), several transaction mechanisms were explored for the blockchain mm-transactions. They all support the ACID and accesses control properties but provide for different levels of privacy protection. The mechanisms were compared in relative terms of effort required by an attacker to subvert the mechanisms, wherein the "relativity" is in terms of whether one mechanism requires more or less effort on the attacker's part in subverting the privacy mechanism than the attacker requires in subverting another privacy mechanism. The costs of the mechanisms were also evaluated.

*Transactional properties for BPMN Trade Transactions*

We adopt the blockchain mm-transaction properties as the transactional properties to be enforced by the system:

– *ACID* properties must be supported.
– *Access control* and *privacy* should be supported.

*3.2.2 Supply-chain Use Case*

To provide examples when discussing transformation of BPMN models into smart contracts, we adopted the supply-chain use-case from (Weber 2016 [55]) that was subsequently also used in (Tran 2018 [51] and [López-Pintado 2019] [34-35]). The use case, shown in Fig. 3 represented as a BPMN model, begins with the buyer issuing a new order. Please, ignore the rectangular boxes labelled as S1, S2, S3, and S4 for now as they represent the result of analysis described in the next subsection. Once the manufacturer receives the order, she/he calculates demand and places an order with a middleman. Middleman concurrently sends the order to the supplier and orders transport from the special carrier. The supplier fabricates the product and prepares it for transport. Carrier, upon receiving the request from the middleman, requests details from the supplier. Supplier provides the details to the carrier and then it prepares and provides the waybill to the carrier. Upon receiving details about the product and the waybill, both from the supplier, the carrier delivers the order to the manufacturer. Upon receiving the order, the manufacturer starts the production and when that is finished, it delivers the product to the buyer, who receives the order.

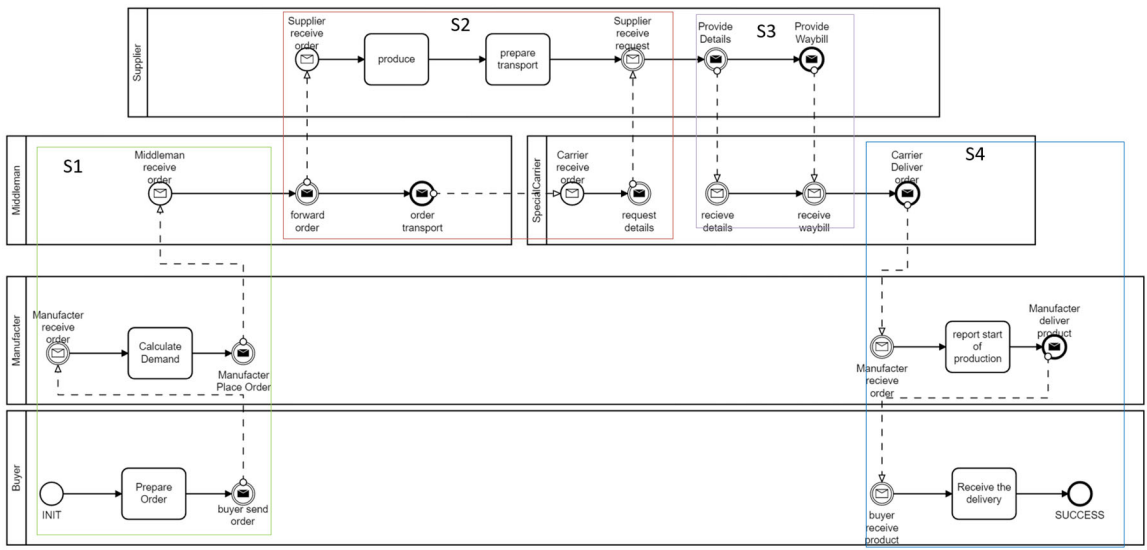

Fig. 3. BPMN model and found subgraphs mapped back to the BPMN model. (Adopted from (Bodorik 2023 [4])).

There are five actors (Buyer, Manufacturer, Middleman, Supplier, and Special Carrier), with each actor having a pool, or a swimlane within a pool, that is represented by a large rectangle representing processing performed by the actor. Arrows, represent connecting objects, which are either sequence flows or message flows. Only message flows may cross pools. It should be noted that some of the BPMNM elements represent tasks that appear in the diagram as smaller rectangles with rounded corners. A task is an activity, which is a script provided by the developer, that is executed when the flow of execution, as represented by arrows and other elements, reaches the task. For each task, the developer provides the code. The remaining BPMN elements are used to represent the coordination/synchronization of the workflow of the trade activities as performed by the actors. Furthermore, when two or more arrows exit out of an element, the element serves as a split/fork/diverging gateway, such that each arrow represents a concurrent stream of activities performed by the actor. When more than one arrow end at (emanate from) a BPMN element, that element serves as a join/merge (split) gateway representing. Split gateways may be inclusive or exclusive, wherein an exclusive gateway allows only one stream to leave, while inclusive gateway may allow many streams to proceed with execution concurrently. An inclusive joining gateway passes a token only if all incoming paths are enabled. Both splitting and merging gateways may have a condition specified on each path.

*3.2.3 How to represent Trade Transaction in BPMN Models*

BPMN modeling specification (BPMN 2.0 2010 [8]) provides the concept of an all-or-nothing transaction only for the subprocess element, which is a BPMN sub-model with one start and one end element. However, as the sub-process is executed within the context of a single actor, it cannot be used to represent the collaborative activities of more than one actor. Here we address the issues of how to determine which BPMN patterns are suitable for representing a trade transaction, and how the developer indicates to the system which BPMN pattern is to be treated by the system as a trade transaction.

Clearly, selecting any subset of BPMN elements from a BPMN model is not appropriate as the pattern selected as a transaction must satisfy the all-or-nothing property during execution. Consider the sample use case, shown in Fig. 3, and the activities of the *Middleman* actor, which are: *Middleman receive order, forward order,* and *order transport*. It may appear that the three activities might be a suitable pattern to be declared as a BPMN transaction; however, difficulties arise with the atomic commitment. If the first two activities are successfully performed but the third one fails, then the transaction is aborted, and the question arises: What to do about the effects of the *forward order* activity that has already been executed, which means that the order was already sent to the *Supplier* actor? The *Supplier* actor might have *produced* the product that is now prepared for transport – and these activities would also have to be aborted and further cascaded aborts might be required. The problem arises because the activities selected as a transaction were not localized in that they affect other activities not belonging to the transaction. This is similar to the *isolation* property of the transactions in the DB system: *No activities external to the transaction should view updates made by a transaction that has not been completed*.

In terms of the graph representation, the three activities that were selected as a transaction are such that there is a flow of execution, from the selected BPMN pattern of the three activities, to an activity outside the pattern. To avoid such a situation, nodes that are selected to be the pattern representing a transaction must be such that there are no outgoing edges, out of any vertices in the pattern, to any vertices outside the pattern. The exception is one vertex that may have an outgoing edge(s) corresponding to the outgoing flow of execution in the BPMN model when the transaction is completed. And the above requirements are satisfied when a pattern chosen to be a transaction is a SESE subgraph. Once the flow of computation enters a SESE subgraph via its *entry node*, it remains within that subgraph until the computation exits via the subgraph's *exit node* – we refer to this as the *localization property* of SESE subgraphs and we exploit it, together with additional properties exhibited by the SESE subgraphs described below, in defining the BPMN trade transactions and in creating the transactional mechanisms to support the BPMN trade transactions.

For the developer to determine which of the BPMN patterns are to be supported as trade transactions, the DAG representation of the BPMN model is analyzed in order to find all SESE subgraphs. Once all such subgraphs are identified, they are mapped back to the BPMN model and presented to the developer who determines which of the subgraphs should be transformed into a BPMN trade transactions. Once the developer decides which of the SESE subgraphs are to be processed as trade transactions, the BPMN model is further processed to create a transactional mechanism to enforce the transactional properties.

Consider the sample use case shown in Fig. 3. The graph is analyzed to find all SESE subgraphs in the DAG representation of the BPMN model. Fig. 3 also shows the SESE subgraphs when mapped backed to the BPMN model – they appear as faint rectangular boxes labeled as S1, S2, S3, and S4. It should be noted that that there are many other SESE subgraphs that exists but are not shown in the Fig. 3, as the figure would become too busy. How SESE subgraphs are further classified and how they are used by the TABS tool will be discussed later.

The subgraphs shown in the figure are:

- S1 … The subgraph contains INIT, Buyer sends offer, Manufacturer receives order, calculate demand, Manufacturer places order and Middleman receives order.
- S2 … The subgraph contains forward order, order transport, supplier receives order, produce, order transport, prepare transport, carrier receive order, supplier receive request and request details.
- S3 … The subgraph contains provide details, provide waybill, receive details, and receive waybill.
- S4 … The subgraph contains carrier deliver order, manufacture receives order, report start of production, manufacturer deliver product, buyer receives product, and SUCCESS.

The developer chooses which of the subgraphs should be used to form trade transactions, one transaction per SESE subgraph. In the next subsection we shall discuss the SESE subgraphs property that, if any two SESE subgraphs share vertices, then one of the subgraphs is a proper subset of the other subgraph. This property is used to define nested trade transaction that will be discussed in the next section. In this subsection 3.2.3, we described how a BPMN trade transaction is identified and specified by the developer, while how a transactional mechanism is created for a smart contract trade transaction is described in the next subsection. How the nested trade transactions are supported is described in the sub-section 3.3.

*3.2.4 Transaction Mechanism to Enforce Trade Transaction Properties*

In this subsection, we describe how the TABS transformations are amended to support the BPMN trade transactions. The main effect is on the modeling at the DE-HSM level that is used to determine which BPMN patterns are suitable for defining as trade transactions. Thus, BPMN model preprocessing and transformation of the BPMN model into the DAG representation are not affected, and we describe here the effects of supporting the trade transactions on DE-HSM modelling and subsequent transformations.

*DAG Transformation into DE-HSM and DE-FMS Sub-models and Properties of SESE Subgraphs*

As was described above, the developer uses the TABS tool to find and show all SESE subgraphs from which the developer chooses which of the subgraphs are to be treated as trade transactions. As was shown in (Bodorik 2023 [4]), the SESE subgraphs satisfy the following properties:

- As the BPMN graph representation is a DAG, each of the SESE subgraphs is a DAG.
- As a SESE subgraph is acyclic, the subgraph may be analyzed using multi-modal modeling in which concurrency is modeled using DE modeling and functionality is expressed using HSMs.

There are two consequences due to the above properties:

– In DE modeling, during execution, each subsystem represented by a subgraph contains a queue of events ordered by their timestamps. As none of the SESE subgraphs have feedback loops (they are acyclic), all DE-HSMs models may use just one DE queue of events ordered by the event timestamps.

– More importantly, recall that the edges in the DAG represent the flow of computation. As SESE subgraphs have only one *entry node*, once computation enters the subgraph via the *entry* node, the computation proceeds with the execution within the subgraph until the execution exists via its *exit node* – hence we refer to a SESE subgraph as representing a *localized computation*.

It is the localization property of the SESE subgraphs that we exploited in (Bodorik 2023 [4]) for automated transformation of an FSM into a smart contract with sidechain processing, wherein the functionality of a selected SESE subgraph is transformed into the methods of a smart contract that is deployed and executed on a sidechain. It is a "slave" contract of a master smart contract that is executed on the mainchain and that invokes the methods of the slave smart contract. As the slave contract executed on a sidechain has localized computation, once the flow of execution enters the subgraph via its *entry node*, it remains within the subgraph until the flow of computation exits via the subgraphs *exit node*. It should be noted that if a transaction is aborted, then its abort compensation activities are performed within the scope of the transaction and exit from the transaction is via the subgraph's exit node.

There are two additional properties that SESE subgraph satisfy (Bodorik 2023 [4]):

1) For any two SESE subgraphs found to exist in a DAG, they are
   - either *mutually exclusive* in that they do not share any vertices, or
   - *one of the subgraphs is a proper subgraph of the other one*.

2) Any vertex s ∈ S, which is neither an *entry* nor an *exit vertex* of the SESE subgraph, has exactly one incoming and one outgoing edge.

It should be noted that the second property is to ensure that analysis of the DAG does not decompose a sequence of n serial activities into n SESE subgraphs, each containing one activity. For more details, please see (Bodorik 2023 [4]).

*Transformation of the DE-HSM Sub-models into Smart Contract Methods*

Recall from the subsection 2.3 that when there is no sidechain processing, then activities of each actor are represented by a smart contract method representing the flow of execution for that actor. The flow of execution for one actor's model is represented using an FSM, if there is one flow of execution, or using concurrent FSMs if there are concurrent executions within the actor context due to the BPMN flow for that actor having fork gates. However, if the developer chooses a transaction representing a collaboration of several actors, i.e., if the developer chooses a SESE subgraph as a trade transaction, our approach represents such a transaction by using a separate flow of execution. Thus, for each SESE subgraph chosen by the developer to be treated as a transaction, a separate smart contract method is created. The flow of execution within that method is represented by the concurrent FSMs that represent the flow of execution within that SESE subgraph chosen by the developer as a trade transaction. Thus, the smart contract is represented by (n+m) smart contract methods, where n is the number of actors and m is the number of BPMN unnested trade transactions as chosen by the developer.

A smart contract method representing a trade transaction does not invoke any of the other non-transaction smart contract methods. Thus, although the smart contract method for a trade transaction is invoked multiple times by an application, the method does satisfy the property that once computation enters the transaction as represented by a SESE subgraph, it stays within that computation until the computation exits the subgraph, which satisfies the *isolation property of an mm-transaction* as described in (Liu (2023) [33]). As a consequence, the methods for automated creation of an mm-blockchain transaction mechanisms are applicable also for the transactional mechanism for trade transactions as is described next.

*Transactional Mechanism for Blockchain Multi-method Transactions*

As stated before, blockchains support the concept of a transaction that is the result of an execution of a single method and do not support a transaction that is the result of executing more than one method execution, whether it is a multiple execution of the same method or the result of executing two different methods of the smart contract. In (Liu 2023 [33]), we describe how to define an mm-transaction as a subset of the methods of a smart contract under the constraint that such transactions methods are *isolated*. The isolation property prevents the transaction methods to make references to objects, or invoke methods, that are declared externally to the mm-transaction methods as that would violate the atomicity and isolation properties of a transaction. Recall that the isolation is violated if a transaction method invokes an external method, as the external method may affect computation that is external to the transaction – and if the transaction is aborted, the activities of the method X would also have to be aborted/negated. As described in (Liu 2023 [33]), implications of providing a transactional mechanism to support a blockchain mm-transaction are:

1) Blockchain is immutable: Thus, to support an mm-transaction, optimistic methods need to be used in support of the transaction commitment and the transactional properties. An mm-transaction method must perform ledger writes in some private workspace and only when the transaction is committed are the writes written from the private workspace to the ledger.

2) The private workspace to support optimistic transaction methods is needed to store the reads and writes to the ledger before the transaction commit. Only when the transaction commits are the cached data applied to the ledger. Furthermore, the private workspace must be shared across executions of the methods of a multi-method transaction.

However, blockchains may not provide a data structure that persists across executions of methods of a smart contract. An example of a persistent data structure provided by a blockchain is *private data* provided by the Hyperledger Fabric blockchain in form of a private blockchain shared by specified actors only. However, if such a data structure is not provided, then there is no alternative but to store the shared information on the blockchain itself, potentially with a support for off-chain storage techniques.

To automatically generate a transaction mechanism, we used the pattern augmentation techniques (Liu 2023 [33]). The developer first creates the smart contract including the methods that are intended to constitute the mm-transaction, wherein the transaction methods must satisfy the isolation property. Once the smart contract methods are created, the developer identifies/marks the transaction methods so that they are known to the preprocessor as methods belonging to the mm-transaction. Before the smart contract methods are compiled, they are examined by a preprocessor that amends each of the transaction method in the way described below. Besides the smart contract methods, the preprocessor is provided with additional information on who the actors are that participate in the transaction and information on how the private workspace is provided, which is discussed later. In short, the preprocessor amends the smart contract methods with patterns as follows:

– The pattern representing the *begin transaction* method is inserted. It initializes the private workspace that must persist across executions of the transaction methods and be accessible to all methods of the transaction. The private workspace acts as a cache to store:

  o The state of the transaction that is managed only by the patterns augmented by the preprocessor. In other words, the transaction methods prepared by the developer are aware neither of the cache nor the object that is used to store the state of the transaction – these are augmented by the preprocessor.

  o The ledger reads and writes performed by the transaction methods.

– Each read or write to the ledger made by any of the transaction methods must be replaced with a pattern to make that read or write using the cache instead.

– End of the transaction method is inserted. It first generates an event, to be processed by each of the transaction participants, that provides information on the results of the transaction so that the transaction participants can review the results and approve them by cryptographically signing them. After the participants approve the transaction results, the method propagates all transaction reads and writes, stored in the cache, to the ledger itself, together with the certifications of the results by participants. As is noted in limitations, we do not as yet implement the reception of the certifications from the participants as a part of the transaction commitment.

The begin and end transaction methods, and the declaration and manipulation of the object representing the state of the transaction, are facilitated by the preprocessor using the pattern augmentation techniques.

(Liu 2023 [33]) provides the following options for providing the transaction workspace/cache:

1) Cache is hosted in the private data structure if it is provided by the native blockchain.

2) The cache is hosted on the ledger in locations that are known only to the methods of the smart contract and hence the cache is "hidden" from other non-transaction methods.

3) Slave smart contract is used to host the cache and the methods of the mm-transaction. The methods are invoked by the main smart contract that contains all non-transaction methods that invoke the transaction methods of the slave smart contract. There are two variations:

  a. The slave smart contract is deployed on the same chain as the main smart contract.

  b. The slave smart contract is deployed on a sidechain, while the main smart contract is deployed on the mainchain.

It is shown in (Liu 2023 [33]) that all the above options support the ACID properties. How augmentation is used to support the properties of access control and privacy is also described. Access control assures that only the actors participating in the transaction can invoke the transaction method. This is achieved by augmenting each transaction method with a pattern that checks that the actor, causing the invocation of the transaction method, is in the list of the actors that may participate in the transaction.

There are difficulties with supporting the privacy property, which states that only the actors that participate in the transaction have access to the cache. Although access control to the methods is supported, the issue is that any actor that has access to deploy smart contracts on the blockchain has access to any data that is on the blockchain ledger. For privacy, (Liu 2023 [33]) examines architectures available to support the transaction as described above while also supporting privacy. For instance, if the transaction methods are deployed in a separate smart contract on a sidechain, privacy is supported as long as the actors who do not participate in the transaction do not have privileges to access the sidechain. Another option for privacy is to use encryption on the private workspace; however, the cost estimates indicate that this may be expensive if public-key cryptography is used to encrypt all data stored in the private workspace. As an alternative, in order to decrease the cost of cryptography, instead of encrypting all data stored in the cache, which is on the ledger, only the location of the cache on the ledger is encrypted: If an attacker does not know where on the ledger the data is stored, the attacker cannot view the stored data. Further details may be found in (Liu 2023 [33]).

### 3.3 Nested Trade Transactions

Consider Fig. 3 that shows the BPMN model of our use case with the four SESE subgraphs identified by the TABS tool. As already noted, there are other SESE subgraphs that contain some other SESE subgraphs. For instance, Fig. 4 shows an additional SESE subgraph, S5, that contains proper subgraphs S1 and S2, wherein each of S1 and S2 are SESE subgraph, representing potential sub-transactions of S5, that do not contain any proper subgraphs that are also SESE subgraphs. In terms of selecting the trade transaction, there are several options available to the developer, such as:

1. Select S1 and/or S2 as separate trade transactions.
2. Select S5 as a trade transaction (but without selecting S1 or S2 as a sub-transaction).
3. Select S5 as a trade transaction and select one or both of S1 and S2 as sub-transactions.

The last option results in nested transactions, which is an issue that we address in this sub-section. We shall concentrate on describing a simple 2-level nesting of a transaction with two sub-transaction and then we comment on how the approach is generalized to an arbitrary number of nesting levels and sub-transactions.

If the user selects only S1 and S2 as two separate trade transactions, then the smart contract will have (n+2) smart contract methods: One method for each of the n actors of the contract, plus one method for each of S1 and S2. For each of the trade transactions S1 and S2, pattern augmentation techniques described in the previous section are used to amend the smart contract methods with patterns to create a transactional mechanism for each of S1 and S2.

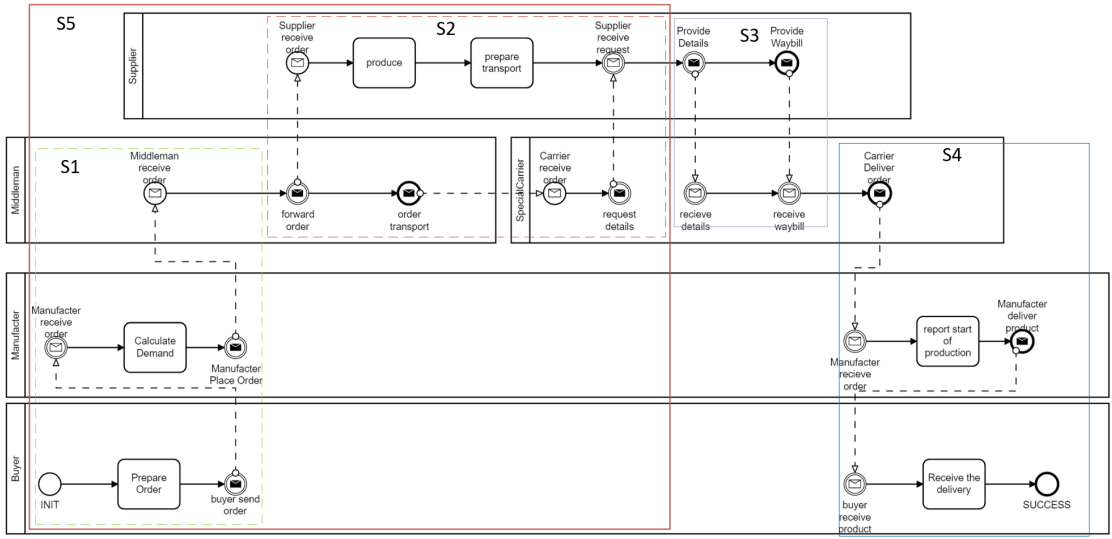

Fig. 4. SESE subgraph S5 contains SESE subgraphs S1 and S2.

When a developer selects a transaction with nested sub-transaction, the construction of the transactional mechanism for the nested transaction proceeds bottoms up, starting with creating the mechanism for the innermost transactions. First, the smart contract methods for the innermost transactions are constructed in the usual manner, one method for each of the innermost trade transactions as described above. If there is a trade transaction with m transactions nested within it, then there will be (n+1+m) smart contract methods: One for each of the actors of the smart contract, one for the outer transaction (the parent) transaction, and one smart contract method for each of the child sub-transactions.

However, atomic commitment needs to be supported for the parent and its child sub-transactions; that is, the commitment of the parent and child transitions together must be atomic. To support the atomicity, we use the 2-Phase-Commit (2PC) protocol, wherein the parent-transaction smart-contract method is augmented with a pattern for the 2PC coordinator of the 2PC protocol, while each of the sub-transactions is augmented with the pattern for the 2PC participant. Thus, before the parent transaction may commit, each of its participating sub-transactions needs to be prepared to commit. Once the parent and children of the 2PC are ready to commit, the parent commits and then child sub-transactions commit.

Recursive application of the above approach yields transactional support for nested transaction to an arbitrary level. As the 2PC protocol is a standard technique that has been used extensively in distributed systems, we shall not elaborate on details, but rather estimate the overhead caused by supporting nested transactions in the next section.

## 4 TABS+: PROOF OF CONCEPT AND COSTS

As a PoC that our approach to supporting nested BPMN transactions is feasible, we amended the previously described TABS prototype tool to also support the trade transactions with nesting and thus create a tool called TABS+. We first describe our experimental evaluation and its environment and show delays to transform the BPMN model of our use case into the methods of smart contracts. We then show how the tool is used to support a trade transaction without nesting and provide cost estimation. We then show how the nested transactions are supported together with cost estimation due to the transactional mechanism to support nested transactions. We also discuss the benefits and drawbacks of supporting nested transactions.

## 4.1  Experimental Evaluation

We performed the evaluation using as setup that included both Ethereum and Hyperledger Fabric (HLF) test chains. The TABS+ tool, designed as a two-tier application, incorporates a NodeJS frontend handling tasks like BPMN diagram composition using BPMN.io, diagram transformation to DE-HSM mode via the GraphViz JS library, and other functionalities through a suite of dedicated JavaScript libraries and APIs for compiling and deploying Smart Contracts on Ethereum and Hyperledger. We used AES for encryption/decryption. Data is encrypted before caching and decrypted for subprocess use, ensuring access is restricted to authorized entities only. Moreover, we combined AES encryption with SHA-256 hashing to validate the integrity of the encrypted data, enhancing security.

For our testing environment, we utilized three cloud servers from DigitalOcean, each provisioned with 2 CPUs, 4GB of RAM, and 80GB of SSD storage, running Ubuntu OS. Ganache-CLI was employed on these servers, emulating a realistic Ethereum mainchain environment with configurations mimicking public network parameters. Sidechain networks were set up as Quorum instances for Ethereum and utilized Microfab for HLF Chaincode.

The servers also function as IPFS nodes within a private cluster, ensuring dedicated and secure data access for each participant in the test network. IPFS was used for off-chain storage, which is a standard approach to minimizing data stored on the blockchain.

The TABS+ tool is configured to support testing by stepping through the transformation process and checking inputs and outputs. It is also setup to enable capturing of timing at various points so that we can measure various delays.

For testing and validation, we used a combination of approaches. To check if our transformation solution works correctly, we tested transformations of BPMN models of different complexities by varying different parameters for the models, such as number of participants, complexity of transactions, and nesting of transaction. For each model, we reviewed and checked the intermediate transformations from BPMN model to the DAG representation, then to DE-HSM model, then to DE-FSM model and finally to the methods of smart contracts. For each transformation we checked the results for correctness and completeness.

As was performed by other approaches, we also tested conformance testing of various traces, wherein a trace is input stream from the Dapp to the smart contract, and conformant trace means that the trace was correctly handled; that is, conformant trace contains a valid sequence of inputs, as described by the BPMN model. As expected, our system achieved 100% conformance in identifying the valid and invalid sequences. This is not surprising as the transformations provide direct mappings from BPMN to intermediate models and finally to the smart contracts methods that use the DE-HSM model with its queue of events for sequencing of those events.

We also carried out conformance testing to check the correctness of cross-chain smart contracts execution. For conforming testing, we compared the results of the transformed smart contracts invocations with the expected outcomes defined in the original BPMN models. We made sure that all the tasks, conditions, and transactional properties were executed correctly as specified.

### 4.2 Transformation Delays

Table 1 shows delays for the transformations of a BPMN model, shown in Fig. 4. Delays to transform the PBMN model into DE-HSM model and identify the SESE are shown in Table 1 in the second row with "Model Analysis". The measured delay is from the point of the start of the BPMN model transformation until the SESE subgraphs are shown to the developer to select those to be deployed as transactions. The remaining delays are self-explanatory. All smart contracts we deployed on an Ethereum test chain.

**Table 1**

| Activity\Delay | Delay in ms |
|---|---|
| Model Analysis | 2,946 |
| Smart Contract Generation | 4,856 |
| Smart Contract Compilation | 6,478 |
| Smart Contract Deployment | 29,786 |

It should be noted that that delays in seconds for all categories with the exception of delays for deployment that are in tens of seconds. However, considering that comparable compilation and deployment would have to be expended even if the developer created the smart contract from scratch, the 10 second or so delay for the transformation process is insignificant in comparison to the time a developer would spend for writing the smart contract from scratch.

### 4.3 Trade Transactions without Nesting

Fig. 5 shows two screenshots of the tool for the use case shown in Fig. 3.

– Fig. 5(a) shows the BPMN model that was created using the TABS tool (TABS simply invokes the Camunda BPMN software editor (Camunda [11]) to support the creation of a BPMN model and store it in as an XLM file (Business Process Execution Language (2023) [9]).

– Fig. 5(b) shows the SESE subgraphs forming the DE-FSM models, i.e., when each subgraph does not have any proper subgraph SESE subgraph, as generated by the TABS+ tool.

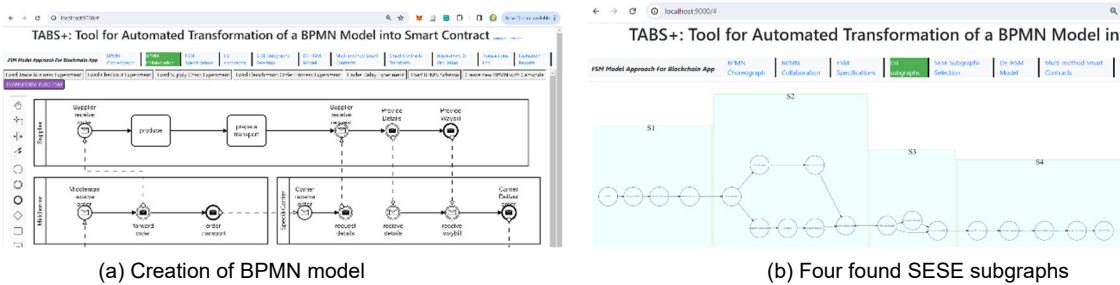

(a) Creation of BPMN model (b) Four found SESE subgraphs

Fig. 5. TABS+ tool use: Creation of a BPMN model finding SESE subgraphs

Recall that the BPMN model goes through a series of transformations resulting in the DAG representation as an interconnection of the SESE subgraphs, wherein each SESE subgraph is represented as a DE-FSM model that may include concurrent FSMs to represent the concurrent streams of executions within a sub-model represented by an SESE subgraph. The interconnection of the SESE subgraphs is shown in Fig. 5(b).

Fig. 6 is a partial screen of the tool showing the SESE subgraphs for selection to be treated as trade transactions. All SESE subgraphs are listed, while the first four subgraphs, S1, S2, S3, and S4, are the four SESE subgraphs shown in Fig. 5(b). The user selects, in the right-hand-side pane, which of the 10 subgraphs, S1, S2, …, S10, are to be treated as trade transactions.

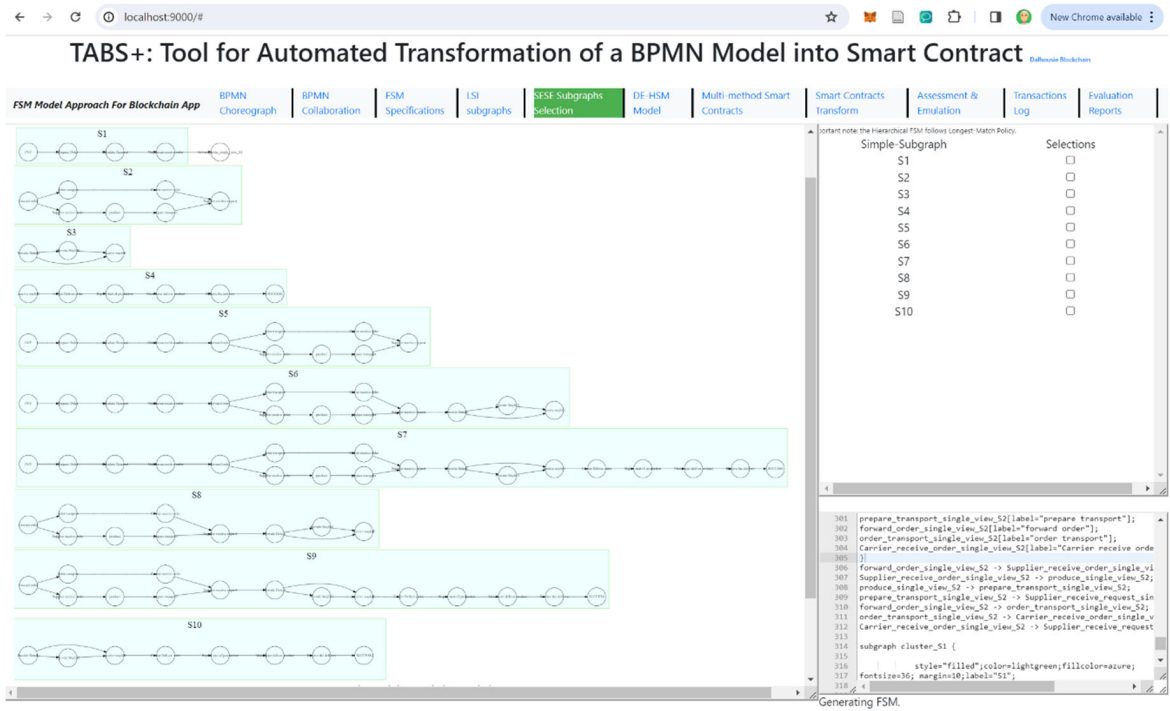

Fig. 6.   TABS+ tool: Listing and selecting SESE subgraphs

In the subsection 3.2.3, we described three architectural alternatives for the trade transaction mechanism, alternatives that depend on the selection of the private workspace to store the transaction's state and the ledger access. We now discuss if and how the TABS+ supports each alternative.

a) *Cache is hosted in the private data structure*, if such a data structure is provided by the native blockchain: We *do not* support this alternative in the TABS+, as it requires specialized treatment, depending on how the private data structure is implemented in the native blockchain. Recall that the private data structure must persist across invocations of the transaction methods and must be sharable by all actors participating in the transaction.

b) *The cache is hosted on the ledger* and all methods are in the same smart contract: As the ledger is accessible to any user who has access to the ledger, an attacker, who has access to the ledger, has access to the smart contract code/script and hence it conceivable for the attacker to deduce form the smart contract methods' scripts where the private workspace is located on the ledger and hence access it.

Consequently, the method will provide less privacy than the other methods for supporting the transactional mechanism that are described below.

c) Slave smart contract is used to host the cache and the methods of the trade transaction: The methods are invoked by the main smart contract that contains all non-transaction methods. There are two variations:

    i.    The slave smart contract is deployed on the same chain as the main smart contract.

    ii.    The slave smart contract is deployed on a sidechain, while the main smart contract is deployed on the mainchain.

As we do not support the first case and the last case had two subcases, the TABS+ tool supports the following three alternatives to achieve the support for BPMN trade transactions:

1. sc-all… One smart contract is used to host all methods, methods or non-transaction methods.

2. sc-2m … Two smart contracts, one is used to host the transaction methods and the cache, while the second one is used to host all non-transaction methods. Both contracts are deployed on one chain, referred to as the mainchain.

3. sc-2s … Two smart contracts, one is used to host all non-transaction methods and is deployed on the mainchain, while the second one, deployed on a sidechain, is used to host all transaction methods.

For the last case, when there are two smart contracts and sidechain processing is used, we used the same cost calculations for both the mainchain and the sidechain. Thus, the estimated cost when there is sidechain processing is higher than when both smart contracts are on the mainchain. As sidechain processing incurs overhead relative to processing on the mainchain only, it is cost-wise advantageous only when sidechain processing is cheap enough that its cost benefits outweigh the overhead costs associated with sidechain processing.

However, as we are interested in determining the costs for all the above alternatives, we configured the tool to provide smart contracts for all three options. To determine the cost of executing patterns for each of the alternative mechanisms, we utilize the Ethereum blockchain tools that enable estimation of the cost of execution of smart contract methods written in the Solidity language. We use the Remix compiler that provides for compilation of Solidity smart contract method into executable code and provides estimates of the total cost of executing a smart contract method by relatively detailed cost calculation for each of the method's instruction. We use the Remix compiler to measure the cost of execution for each method of the smart contract for each of the alternative transaction mechanism approaches and for the base case (labeled as *no-xa*) that does not support trade transactions. To calculate the cost of the mechanisms, we performed measurements using a smart contract that contains the following methods:

- Smart contract *method m (inputParameter X: objectX)*. It receives a parameter that is an object of a specified size, and it invokes the methods m1 and m2 in that order.

- method *m1(x1: objectX)* … the method writes to the ledger the object passed as an input parameter.

- method *m2(x1: objectX, x2: objectX)* … the method reads from the ledger the object written by the method m1 and then writes it back to the ledger again as a new object.

We use the Remix compiler to measure the gas-cost of execution for each method of the smart contract and thus estimate the relative overhead cost of providing a transaction mechanism when compared to the base case when execution is without a transaction mechanism. Table 2 and Fig. 7 show the cost for each of the alternative mechanism in the numerical and graphical representation, respectively. It's important to mention that we've standardized the gas price at 20 Gwei to maintain consistency with the default gas price of Ganache, which we used for cost evaluation in our preceding research papers (Liu 2022 [32], Bodorik 2023 [4]). The first column of the table contains the label identifying the transaction mechanism. The subsequent columns show the cost estimates for the alternative mechanisms when the trade transaction reads and writes a ledger object of size X = 75, 512, 1024, and 1875 KB. Note that when the experiments were performed,1875 KB was the maximum block size (Ethereum's Average Block Size [17]).

The figure and the table show that the cost is indeed directly proportional to the size of the ledger data read/written by each of the approaches. The cost of the transaction mechanism for this particular transaction doubles the cost in comparison to not having transactions. This is not surprising as any ledger data that is read or written is stored in the cache first, and upon the transaction commit the ledger access is replayed from the cache to the ledger itself. Considering that the cache is also on the ledger, the access to the ledger access in essence doubles and hence the cost, which is directly proportional to the leger reads and writes, also doubles as the table and the figure indicate.

All of the alternative mechanisms we described enforce the ACID properties of transactions as defined in Section 3.2 and also enforce access control. Although each one also supports privacy, the level of privacy protection differs as different levels of efforts are required on the part of the attacker to subvert privacy. We assume that the attacker has access to the blockchain and hence is able to read any object stored on the ledger as long as the attacker knows where on the ledger it is stored – thus privacy may be subverted for the case when all methods and the private workspace are all hosted in the same smart contract (case labelled as *sc-all*).

Table 2: CPU Processing cost estimates (in Gwei)

| Label | 7KB | 512 KB | 1024 KB | 1875 KB |
|-------|-----|--------|---------|---------|
| *no-xa* | 4545000 | 31027200 | 62054400 | 113625000 |
| *sc-all* | 9090582 | 62054982 | 124109382 | 227250582 |
| *sc-2m* | 9319500 | 63621120 | 124113000 | 227254200 |
| *Sc-2s* | 9405000 | 64204800 | 127242240 | 232987500 |

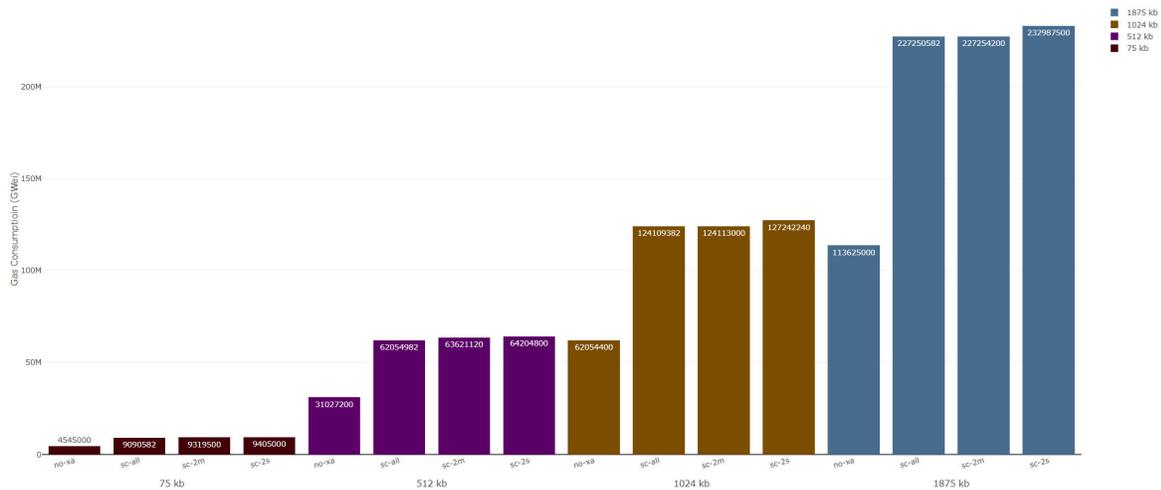

Fig. 7.   Estimated CPU processing costs (in Gwei) as a function of  data size

For the case labelled as *cs-2m*, when the transaction methods and the private workspace are hosted in a separate smart contract hosted on the mainchain, the same approach may be used to subvert the privacy as in the case labelled *sc-all*, except that it would have to be applied twice, once on the main smart contract and once on the smart contract that contains the transaction methods and the private workspace.

The case labelled *sc-2s,* when the transaction methods and the private workspace on hosted in the separate smart contract executed on a sidechain, is similar the case labelled *cs-2m,* except that the attacker also needs access to the sidechain where the smart contract is deployed in addition to having access to the mainchain. Thus, additional information and effort are required in comparison to the previous cases.

To further increase the level of privacy, cryptography may be used to ensure the ledger data written by the transaction is not viewable by the non-transaction actors. To measure the cost of encryption/decryption, the methods m1 and m2, which read and write the ledger, are amended with the public-key cryptography used to encrypt/decrypt any ledger read/write. Thus, the cost of encryption/decryption is expected to be directly proportional to the size of the data that is encrypted/decrypted and thus being expensive, as is the case for the code written in the Solidity language for the EVM [Ethereum Virtual Machine (EVM) [18]). Table 3 and Fig. 8 show the cost of cryptography to encrypt/decrypt the written/read ledger data that has the size X = 75, 512, 1024, and 1875 KB.

Table 3: CPU processing cost estimates (in Gwei)

| Label | 7KB | 512 KB | 1024 KB | 1875 KB |
|---|---|---|---|---|
| no-xa | 4545000 | 31027200 | 62054400 | 113625000 |
| sc-2s | 9405000 | 64204800 | 127242240 | 232987500 |
| sc-2s-crypto | 18268952 | 124706420 | 249411188 | 456684152 |

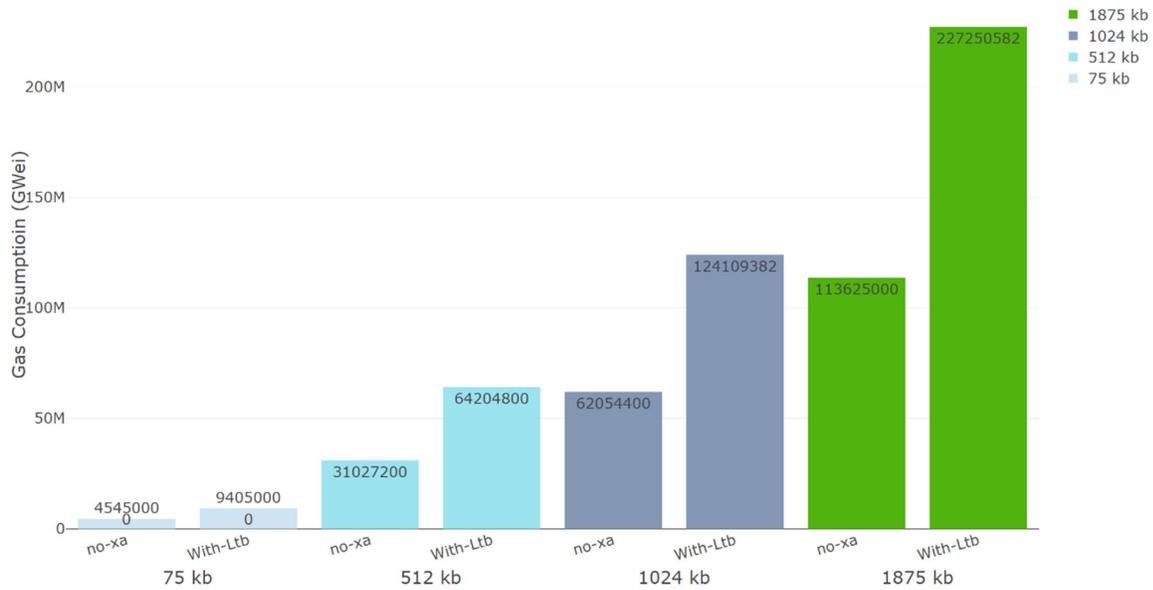

Fig. 8.  *Estimated CPU processing costs (in Gwei) for cases labeled as "no-xa", "sc-2s", and "sc-2s-crypto"*

The table and the figure show the cost of the work performed when: (i) there is no transaction mechanism, labeled as "no-xa" in the first column of the table; (ii) the transactional mechanism is supported using sidechain processing for the sub-transactions, labeled as "sc-2s"; and (iii) the transactional mechanism is supported using sidechain processing for the sub-transactions and when the data stored in the private workspace/cache is encrypted/decrypted when writing or reading, labeled as *"sc-2s-crypto"*. Clearly, the cost of encryption/decryption is significant as it doubles the cost of the transactional mechanism when sidechain processing is used (case labeled *"sc-2s-crypto"*).

### 4.4 Supporting Trade Transactions with Nesting

Recall that steps performed in transformations of a BPMN model leading to and finding the SESE subgraphs are the same, whether the tool is used to support sidechain processing or for supporting BPMN trade transactions. However, once the SESE subgraphs are shown to the user, as in Fig. 6, further transformation and processing to support the trade transaction differs from the case when the TABS tool is used to support sidechain processing. To support a trade transaction, the developer is asked to select those SESE subgraphs that should be treated as a BPMN trade transaction. For instance, the developer may select the subgraphs S3, S4 and S5 as trade transactions. S3 and S4 are SESE subgraphs that do not contain other SESE subgraphs. However, the subgraph S5 contains the SESE subgraphs S2 and S3 as its subgraphs and hence they are shown to the user, as is shown in Fig. 9. They may lead to nested transactions in that S5 contains sub-transactions S1 and S2. The developer may then choose any one of S1 and S2, or both, or neither as sub-transactions. Following this, the developer is provided with options for the selection of the type of the trade transaction mechanism to be used to support the selected nested trade transactions.

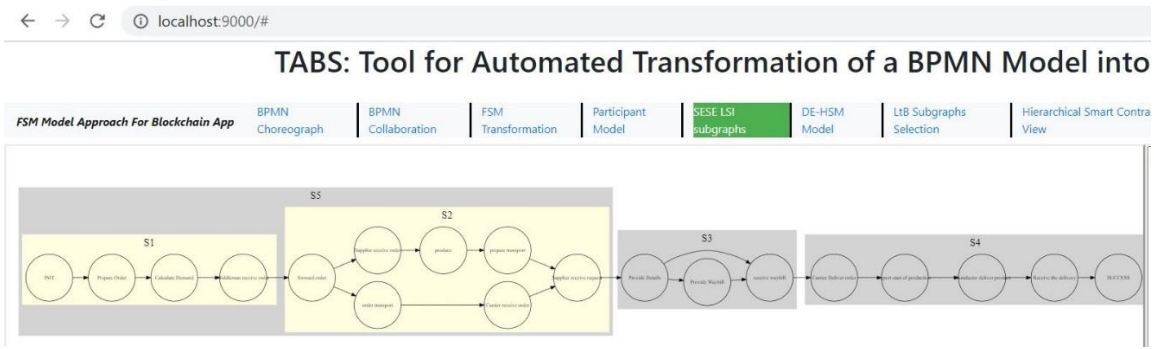

Fig. 9.   User selecting subgraphs S3, S4, and S5, wherein S5 contins SESE subgraphs S1 and S2.

The graphical representation of the methods for the smart contracts are shown in Fig. 10. For each of the trade transactions S3, S4, and S5, there is a method that is denoted by three icons labelled "method_start_SX", "tx_SX", and "method_end_SX", where X is either 3, 4, or 5, depending on the transaction number. However, as S5 has two sub-transactions S1 and S2, it further contains two methods, one each for the sub-transactions S1 and S2, respectively. The figure's layout was produced using Graphviz (Graphviz [21]), while the visual styling was implemented with D3.js (Bostock [5]); both Graphviz and D3.js function as libraries within the Node.js framework (Node.js [43]).

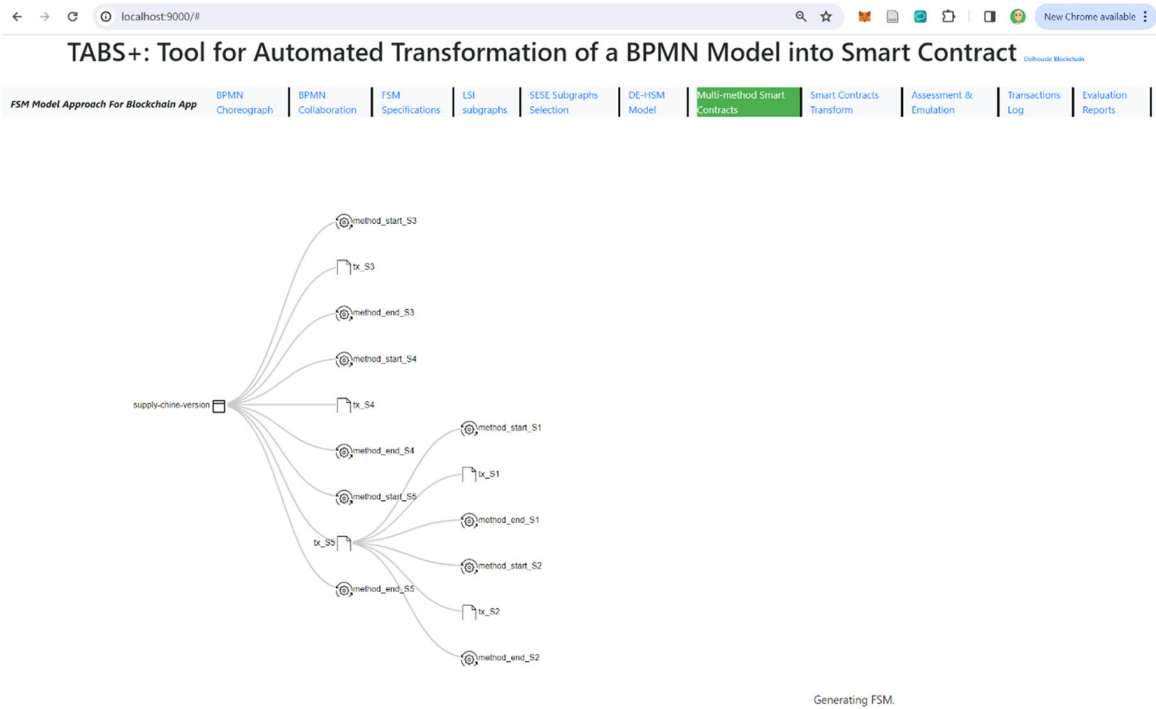

Fig. 10. Methods for trade transaction S4, S4, and S5, whrein S5 has sub-trnasactions S1 and S2.

Recall that to support the atomicity of the parent transaction with its sub-transactions, a 2PC protocol is used, with each parent transaction acting as a coordinator of the 2PC protocol, while each child sub-transaction acts as a 2-PC participant. Thus, S5 includes the functionality of a 2PC coordinator for its sub-transaction S1 and S2, while each of S1 and S2 contain the functionality for the 2PC of a participant. It should be noted that, as a transaction may be a parent transaction and a child transaction at the same time, it may include the functionality for acting as a coordinator for its sub-transactions and as a participant when it is a sub-transaction within a its parent transaction.

The cost estimates for a transaction mechanism to support sub-transactions are expected to be high. In comparison to the cost of a transaction mechanism for a trade transaction that is not nested, for each nested transaction a 2PC is implemented. The cost is incurred by a parent transaction coordinator exchanging the 2PC protocol messages with its participants. Fig. 13 shows the total cost estimates for performing the 2-PC protocol that consists of the coordinator issuing the Prepare-to-commit command and receiving a response from each of the participants, and then issuing the Commit-command to each of the participants. It also includes the cost for each of the participants to receive the Prepare-to-commit command and to send the Ready-to-commit responses to the coordinator and then receiving the Commit-command from the coordinator (the participant does not acknowledge the received Commit-command response). The cost of the 2PC-commit protocol coordination needs to be added to the rest of the cost of the transactional mechanism.

To estimate the cost of the 2PC protocol implementation as a part of the mechanism to support the nested transaction, we evaluated the cost of each step of the protocol, aligning it with the workflow displayed as a sequence diagram in Fig. 11.

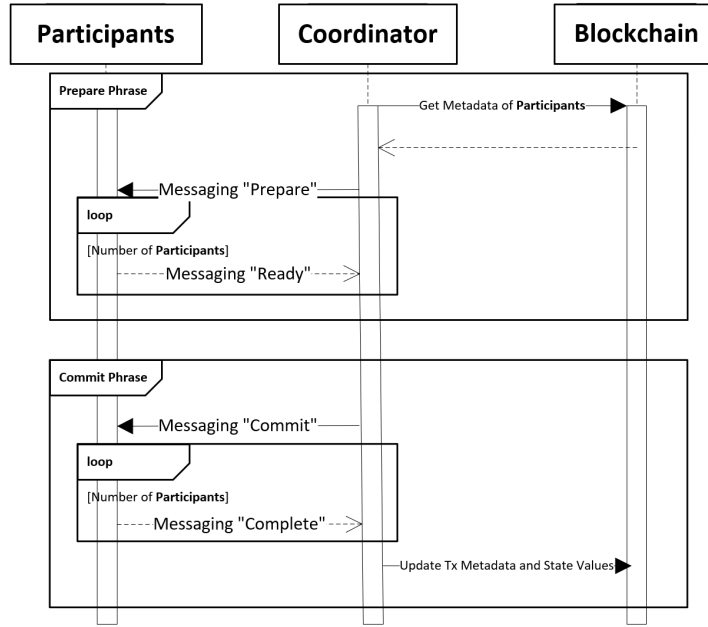

Fig. 11. Sequence Diagram Illustrating the 2-Phase Commit Protocol for Nested Transactions.

During the commitment stage of nested transactions within the parent transaction, the coordinator initiates the 2-phase commit by triggering the 'event' of the smart contracts, which sends a 'ready' message to all its 2-PC participants for the subordinate nested transactions. The execution of this event-emitting function incurs an additional cost. Subsequently, the coordinator awaits responses from the participants through its 'listener' interface. Each subordinate responds by sending back a 'yes' or 'no' message, requiring the response-collecting function to be activated 'N' times, where 'N' equals the number of participants. This is where the overhead of the 2-phase commit protocol is most evident. Similarly, during the second phase, the coordinator disseminates a 'commit' message to all participants, who subsequently respond with their replies. The coordinator declares a successful commitment if all participants return the positive "ready-to-commit" response. The additional cost of the second phase commit is roughly equivalent to that of the first phase.

To gauge the cost of our 2-phase commitment protocol, we employed Remix for interacting with the actual functions of deployed smart contracts. As before, the gas price was set at 20 GWei, identical to the default gas price of Ganache. We estimated the costs of both phases when the number of participants involved in the 2-phase commit varies from 1 to 5, with the results presented in Table 4 for each of the phases of the 2-PC protocol. Fig. 12 shows the total cost of the 2-PC, derived as the sum of the costs shown in Table 4 for Phase1 and Phase2, as a function of the number of participants. As expected, the cost is in a form of a linear function of the number of participants. Furthermore, the cost of a mechanism to support sub-transactions is high due to the 2-PC protocol costs. The cost estimates shown in Fig. 12 need to be added to the cost of the transactional mechanism when nesting is not used, and when they are added, then the cost of the mechanism is more than doubled when there are two nested sub-transactions and increases linearly with an increase to the number of participants.

Table 4: Costs Estimates for the 2-Phase Commit Protocol (in Gwei)

| # of participants | Phase 1 | Phase 2 |
|---|---|---|
| 2 | 627820 | 627400 |
| 3 | 671500 | 671080 |
| 4 | 715180 | 714760 |
| 5 | 758860 | 758440 |
| 6 | 802540 | 802120 |

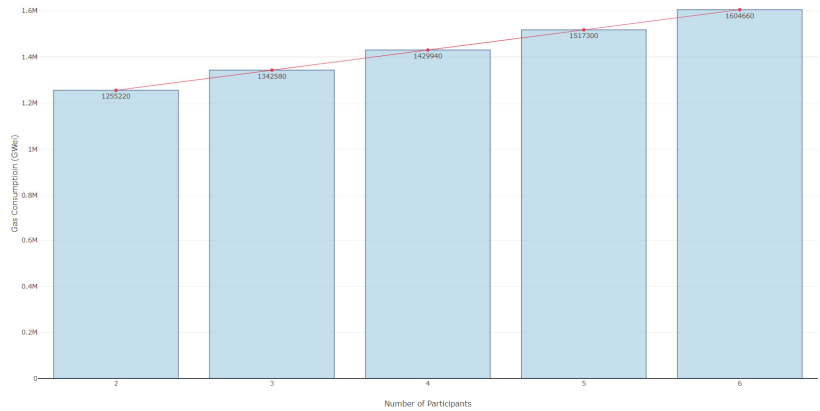

Fig. 12. Gas Utilization by the 2-Phase Commit Protocol for Nested Transactions.

## 4.5  Benefits and Drawbacks of Supporting Trade Transactions with Nesting

### 4.5.1 Benefits

Similar to supporting nested transactions in DB systems, there are numerous benefits gained if their support is provided:

- *Atomicity Across Levels*: All sub-transactions commit or abort as a part of the parent transaction, maintaining atomic integrity.
- *Structured Coordination*: Coordination of activities by the parent and child transactions is clearly defined by interactions within the transaction hierarchy.
- *ACID Compliance*: The ACID transactional properties are correctly enforced in face of complex scenarios.
- Clear Commitment Sequencing: The correct commitment processes is automatically facilitated across multiple levels with the outer transaction as a coordinator.
- *Correctness*: If the automated generation of the transaction mechanism in the transformation phase of a BPMN model into a smart contract is correct, then it is assured that support for nested transactions provided by the TABS+ tool is correct.

- *Developer effort*: If the sequencing of the recovery activities is provided by the mechanism, then the developer's effort to deal with recovery due to exception is eased as the developer needs to provide for recovery/compensation of individual tasks, but not for sequencing of the recovery processes.

*4.5.2 Drawbacks*

- Performance Overhead: As the experimental evaluation showed, the performance hit for using nested transactions is high.
- *Potential Complexity of Handling Failures*: If the transaction mechanism does not facilitate automatic synchronization of recovery activities, then the whole recovery process needs to be provided by the developer.

As blockchain data is immutable, handling failures in blockchain applications is a big headache, and handling failures for nested transactions is no exceptions. However, as our transactions mechanism is based on an optimistic scheme, in which data is written to the blockchain in the commit phase, we believe that an automatic recovery of trade transactions is feasible, and we are currently researching this problem.

## 5  RELATED WORK, FUTURE WORK, AND CONCLUDING REMARKS

We first provide related work, in which we also compare our approach with other contemporary approaches on transforming BPMN models into smart contacts. We then describe limitations of our work and our plans on resolving them. Finally, we describe our future work for the project and provide concluding remarks.

### 5.1  Related Work

Closest to our work is research on transforming BPMN models into smart contracts. One of the earliest works is by Weber et al. (Weber 2016 [55]) that has two transformation phases, which is a standard approach, in that in the first phase the BPMN model is analyzed and transformed into the methods of a smart contract that are deployed and executed on a blockchain, wherein the chosen blockchain was Ethereum. In addition, an off-chain component is used for communication with the Dapp. The actors of the BPMN model exchange messages as specified by the BPMN model, wherein any message exchanged between actors is provided to the off-chain component. The smart contract contains a monitor that stores the model choreography, which is the sequencing of messages as exchanged by the actors who are participating in the Dapp execution. The monitor receives from the Dapp messages that are exchanged between actors and ensures that the messages adhere to sequence specified by the choreography contained in, and enforced by, the monitor. It should be noted that the usage of the term "choreography" in (Weber et al. (2016) [55]) is not to be confused with the "choreography" term as defined in BPMN terminology, in which choreography refers to conversations amongst the actors, wherein a choreography represents an exchange of messages between actors for a specific purpose, but without providing details on the exchange of messages within the choreography.

The Lorikeet project (Tran [51]) builds upon the work of Weber et al. (2016) [55] by augmenting the architecture with a support of business process for asset control, wherein assets include both fungible assets, such as cryptocurrencies, and non-fungible assets (Tran [51]). It supports a registry of tokens and provides appropriate methods for the management of assets, such as transfer of an asset. It claims to support rapid prototyping as smart contracts dealing with asset management can be built readily from a BPMN model which includes support for the asset registry. Thus, smart contracts can be created, tested, and adapted/modified before they are deployed for production.

Caterpillar (López-Pintado (2019) [34-35]) took another approach to transforming a BPMN model to smart contract methods. Unlike the approaches described above (Weber (2016) [55]) and Lorikeet (Tran [51]), which take the collaborative approach of actors as represented in BPMN by a pool, or a swimlane, if actors are within the same organization, all business process are within a single pool at a BPMN model, and the state of all processes and subprocesses are recorded on the blockchain. The architecture contains three layers consisting of (i) a Web Portal, (ii) Off-chain Run-time, and (iii) On-chain Runtime. On-chain Run-time contains a set of smart contracts with various functionalities, such as: the control of workflow, Service Bridge and Worklist Handles to manage interactions with external applications and validate produced results; Contract Factories for configurations and process management; and Log for communicating information to the Off-chain component via events. As in (Weber (2016) [55] and Tran [51]), Caterpillar also used Ethereum as its blockchain of choice, although both permissioned and non-permissioned versions of Ethereum can be used.

Loukil et al. (2021) [36] propose **CoBuP**, a **Co**llaborative **Bu**siness **P**rocess execution architecture, that is based on a blockchain. They also start with a BPMN model, but one of their requirements is to address the issue of a smart contract immutability and hence requiring creation and deployment of a new contract any time new updates are required. Thus, instead of generating a compiled version of a smart contract from a BPMN model, they advocate creation and deployment of one generic smart contract that invokes predefined functions. They propose a system architecture that has three layers: Conceptual, Data, and Flow layer. The conceptual layer provides transformation of the BPMN model, expressed in XML format, into a JSON Workflow model. The data layer contains definitions of the statically encoded data structure stored on the blockchains. The smart contract interpreter for the Workflow model exercises the flow control by invoking process instances that are invoked by the Dapp and that access and update data structures in the data layer model. It should be noted that all participants communicate and interact with the process instance and all activities are executed by the same methods without any support of access control or privacy.

Although Mavridou and Lazska (2018a [38]; 2018b [39]) do not perform transformation of a BPMN model into a smart contract, we need to refer to it as we base our approach to securing the smart contract methods on their work or securing smart contracts derived for a model represented using an FSM. Mavridou and Lazska (2018a [38]; 2018b [39]) start with an FSM representing the application's functionality that they transform into the methods of a smart contract. Then, each method of the smart contract is analyzed and patched for any security holes. They develop a tool, called FSolidM, as a PoC that examines each smart contract method and augments it with security code to eliminate discovered security issues. Furthermore, after hardening a smart contract method, steps are taken to prevent the developer from modifying the inserted security code when amending it before its deployment.

Di Ciccio et al. (2019) [15] compared Lorikeet and Caterpillar approaches, both of which transform a BPMN model into a smart contract, in terms of the following features.

- ▪ *Model Execution*: This feature identifies the *M*odel *Dr*iven *E*ngineering (*MDE*) model, which in all cases discussed here is BPMN. For the execution, it captures if the smart contract on the blockchain controls the flow of execution by generating the code to be executed, or it introduces another layer of abstraction in that the workflow is controlled by an interpreter that invokes processes/tasks to be executed by or on behalf of actors.

- *Coverage of BPMN elements*: BPMN models are built using a set of BPMN elements as described in the background section. As BPMN contains a rich set of elements from which to build BPMN models, all approaches support a subset of BPMN elements and *Coverage* represents the size of the supported subset.
- *Discovery of Incorrect Behavior and Sequence Enforcement*: If inputs provided by actors are incorrect or appear in incorrect sequence, is it handled properly?
- *Participant Selection*: Are participant identities supported?
- *Access Control*: Is access control supported?
- *Asset Control*: Are assets controlled?

We shall compare our TABS+ approach with Lorikeet, Caterpillar, and CoBuP approaches using the above categories as a guidance. We note that the features of *Access Control* and *Asset Control* refer to fungible and non-fungible tokens supported directly by the Lorikeet approach through the transformation process from BPMN model to the smart contract. The access and asset control supports are hardcoded in the smart contract and are used to control access to the registry of tokens to determine if access to the asset is allowed (*Access Control*) or if changes to the asset are allowed (*Asset Control*). Of course, Caterpillar, CoBuP and TABS+ approaches can also support both fungible and non-fungible tokens provided that BPMN models to support them are developed and transformed to smart contracts.

A category that is missing in the (Di Ciccio (2019) [15]) is privacy. Generally, a blockchain does not support privacy, in that any user having access to the blockchain can see anything stored on the blockchain; of course, only if the user knows where to look on the blockchain. From a practical point of view, *privacy* in a blockchain is the ability to keep a transaction state private so that actors/users who are not participating in the transaction are unable to view the details about its state. Consider a trade activity that involves several participants and has several transactions in which different actors participate, such as for the purposes of transport, insurance, or customs clearing. Clearly, if a transaction involves three participants, these participants should have access to the state of that transaction, but the other participants should not have access to the state of that transaction. Thus, we introduce *transaction privacy*, or *privacy* for short, to represent such privacy provision and use it as another feature in the comparison to approaches.

Table 5 provides the comparison. However, our TABS+ approach and its tool support the following additional features that are not supported by the other approaches:

– Support of trade transactions that are supported using mm-transactions.
– Support of nested transactions.
– Deployment of smart contracts on either Ethereum or HLF blockchains. Other blockchains can be supported if the interpreter is written for that blockchain.
– Our approach supports Sidechain and Cross-chain processing.
– Synchronization of activities is achieved in a blockchain-agnostic way as it is represented by a DE-FSM model that is blockchain agnostic. However, the developer still needs to write code, but it is for the independent and isolated tasks with well-defined input and output parameters.
– And, as the table indicates, our approach also supports privacy (as indicated in the table).

Table 5 Comparison of Approaches to Transform a BPMN

| Feature\Approach | TABS+ | Lorikeet | Caterpillar | CoBuP |
|---|---|---|---|---|
| Execution | Interpreter | Code generation | Code generation | Interpreter |
| BPMN coverage | High | Medium | High | Not Available |
| ID Inc. Behavior | Supported | Supported | Supported | Supported |
| Sequence enforcement | Supported | Supported | Supported | Supported |
| Participant Select | Identity based | Predefined | N/A | Predefined |
| Access Control | To SC* methods | Direct support | Via SC | Via SC |
| Asset Control | To SC Asset methods | Direct support | Via SC | Via SC |
| Privacy | transaction based | Registry access | Not supported | Not supported |

*SC … Smart Contract

Below we briefly overview how our previous work on blockchain smart contracts relates to this paper:

o    In (Bodorik 2021 [3] and Liu 2021b [30]) we describe how an application, described using an FSM, can be transformed automatically into the methods of smart contract such that sidechain processing is supported. We describe the transformation process to derive the methods of the smart contract and the system architecture that includes a bridge between the mainchain and the sidechain to support the sidechain processing.

o    In (Liu 2022 [32]) we describe our early work on using DE-HSM models for multi-modal modeling when transforming a BPMN model to the methods of a smart contract.

o    Bodorik 2023 et al. (2023) [4] formalize the approach of the previous work to transforming an application expressed as a BPMN model into the methods of a smart contract. The results of (Liu 2021b [30], Liu 2022 [32]) are used in formulating the transformations in design phase, while we use the results of [Bodorik 2021] in supporting sidechain processing.

o    Liu et al. (2022) [32] raises the issue of how to represent and support BPMN trade transactions.

o    Finally, Liu et al. (2023) [33] describe a mechanism for a developer to specify blockchain transactions that span executions of multiple methods of a smart contract and thus extend the native blockchain mechanism that supports a transaction as a result of executing a single smart contract method. In addition, the paper also describes how pattern augmentation technique is used to automatically create a transactional mechanism for the multi-method transactions specified by the developer.

## 5.2  Limitations and Plans for Their Resolutions

Although we feel that our approach to ease the developer's task in creating smart contracts for trade transactions is feasible, there are still many problems and limitations that need to be addressed. As limitations need to be overcome, we also hint at how we are working on resolving them, and hence in this subsection we describe both, the limitations and our plans on how to address them.

### _Certifications of Transactions_

Certification of transaction results is a relatively straight-forward idea. Once the transaction activities are completed, each of the transaction participants is provided with the transaction results that each one reviews and certifies by signing the transaction results as being correct. How the certification is obtained is described in section 3.2.4. In the transaction commitment phase, the current version of the TABS+ tool does generate an event that provides participants with the information on the result of the transaction that participants are to sign and return to the commitment process; however, the tool does not yet provide for the collection of signed certifications. We are augmenting the tool to support the the collection of the signed certifications.

Even though the transaction activities are executed by the blockchain and hence are deemed secure, different participants provide different inputs to the transaction based on their understanding of the expected activities. The certification steps provide the participant with the opportunity to ensure that the transaction results comply with the participant's expectations and thus aid in guarding against incorrect implementation of the transaction activities either due to errors or misunderstanding of the transaction semantics by the developer when writing the script for the individual tasks.

### Securing Smart Contract Methods

Another extension is to provide for automated hardening of the smart contract methods by automatically ensuring that the security best practices for the creation of the smart contract methods are followed. As an approach to securing smart contracts, we adopted the approach in (Mavridou and Lazska [38-39]), wherein they propose hardening of smart contracts created by transformation of an FSM to smart contract methods. Given an FSM as a representation of the smart contract activities, they proposed a transformation of the FSM into the methods of a smart contract. They then propose securing each of the smart contract methods by inserting security patterns to guard it against (i) reentrancy by using locking, (ii) transaction ordering in face of unpredictable states, (iii) timed transitions, and (iv) access control. We successfully incorporated into the TABS+ the reentrancy protection by inserting appropriate locking patterns into the smart contract and support access control. In essence, the security patterns are inserted into the smart contract method at the start of a method and at its end. Admittedly, we only ascertained that the Mavridou and Lazska's approach [38-39] can be applied to our generated smart contracts, but in our future work we shall develop smart contract patterns to guard against the known smart contract vulnerabilities.

### Blockchain Agnostic Smart Contracts

One of our objectives is that the generation of smart contracts should be blockchain agnostic. We made progress towards this objective as the collaboration is blockchain independent as it is expressed in terms of interconnection of the DE-FSM models, and it uses DEs to model concurrency and concurrent FSMs to model the functionality. However, currently, to apply a smart contract developed for one blockchain to be deployable and executable on another blockchain, the scripts need to be provided by the developer for the task elements to be executable on the target blockchain. To overcome that issue, we are currently applying the two-layer approach taken by the Plasma project, described in (Buterin (2015 [10]), in which the task scripts are not executed on the blockchain, but rather off-chain, while the smart contract simply guides the collaborations and obtains certifications about the results of the tasks that are executed off chain.

### Recovery Process

One of the most demanding and important aspects in software development deals with recovery when something goes wrong. Recovery in blockchains is further complicated due to the immutability properties that force usage of optimistic approaches to commitment, such as the use of 2PC for commitment of distributed transactions. The developer not only needs to write recovery for each of the activities but also needs to worry about their proper sequencing and thus, possibly introduce errors. Using information on nesting of transactions provides the opportunity for the TABS+ tool to automatically facilitate sequencing of such recovery according to the nested structure of the transactions. This would not only ease the developer's task in writing recovery procedures, but it would also reduce potential errors in introduced by the developer when sequencing of the recovery activities.

### Validation and Verification

Although we perform validation and verification of the smart contract produced for exposition purposes, validation and verification needs to be an integral part of the transformation process from BPMN models to smart contracts deployment. We plan to incorporate such procedures for the results of transformation into DE-HSM model, in order to ensure that the DE-HSM model transformation is correct, and for the transformation of the DE-FMS model into the methods of smart contracts. In addition, validation and verification of the TABS+ interpreter, which controls the execution of the discrete events as they arise, needs to be performed.

### 5.3  Future Work

Thus far, we concentrated on the design and development of the transformation process and all aspects that lead to creation of a tool as a PoC that our approach is feasible. Issues still to be addressed, and on which we are already working, were already described above under limitations, together with a brief outline on our approach to overcome them. Our tool is thus organized by the needs of design and development, testing, and validation. The tool has functionality to step through each individual activity for testing purposes to examine partial results. Below we describe our near-future work focusing on augmenting the TABS+ tool for testing acceptance of the tool by developers, which is followed by our long-term plans describing how we further plan on creating adaptation of BPMN models to satisfy constraints or requirements of a particular context for the blockchain application, such as requirements due to geographical location or safety requirements. Furthermore, we describe how we plan on to facilitate to support the relatively new concept of SC-as-a-Service (SCaas).

### 5.3.1 Acceptance by Developers

Although acceptance of our approach and the TABS+ tool by developers is the main objective, the current state of the TABS+ tool is not yet ready for evaluation by developers not only due to the problems described in limitation, but also due to the nature of the tool and its current interface targeted to the design and testing; the tool is not production-type tool developers would expect. In general, we find it difficult to attract developers to expend, without compensation, their time and effort to test a new tool. Thus, we are not yet at a stage to do formal experiment on the tool's usability and its acceptance by developers. However, we did show the tool to some developers in a blockchain company with which we associate and received favourable comments on the approach and the tool, and we also received two suggestions for improvement described below. Of course, these would be in addition to overcoming the limitations described in a previous subsection and improving the user interface (UI) according to the UI design and user experience (UX) guidelines.

### Querying the State of the Smart Contract and Smart Contract Dashboard.

Being able to query the state of the smart contract is one of the features used in (DiCiccio 2019 [15]). We are close to providing the feature as the TABS+ tool has the ability to step through the smart contract activities as inputs from the participants arrive and observe the state changes due to each input. We plan to augment the transformation to also create a smart contract method that returns the current state of the smart contract.

However, we still need to facilitate access control to such information as only the participants to the (nested) transaction should be able to view the state of that transaction. Furthermore, presentation of the dashboard to the user requires significant software development effort.

### *BPMN Model Information for Task Scripting*

Another important suggestion we received is to augment the BPMN model with information for each of the task to include the purpose of the task and, in particular, information on the input and output parameters for the task. Recall that a task in a BPMN model is an isolated fragment that has one entry and one exit. Furthermore, the task data is isolated as the task can only refer to input and output parameters and data that declared are within the task itself. To ease the developer's task in understanding semantics to be provided by the task, when creating the BPMN model, the task purpose and description of its input and output parameters should be described using the BPMN *data association and sequence flows* elements. When the developer needs to write the task code, information about the task, purpose and input and output parameters, should be automatically retrieved from the BPMN model and provided to the developer.

### *5.3.2 Future Concepts Exploiting the TABS+ Approach*

Assuming that all goes well in addressing the limitations and that the approach and the tool are acceptable by developers, we plan on exploiting the results to provide further assistance to developers in creating and managing smart contracts for trade transactions.

### *Repository of BPMN Patterns*

We shall investigate approaches to augmenting BPMN models with patterns or approaches to replace certain patterns with other similar BPMN patterns. The objective is to support a repository of patterns classified by their functionality and data flow, including inputs to the pattern, outputs from the pattern, and methods used. In such a repository, a BPMN model is simply a pattern. For instance, a BPMN model/pattern can represent collaboration of activities in a letter of credit that was described in the Motivation section 1.1. Recall that the model, in addition to the seller, buyer, and banks, other actors may be involved in order to provide for transport of the product to the port, wherein the transport may include: specific safety requirements and insurance, customs clearance processes including documentation required when crossing borders, release of the goods from the port, transport to the seller, and other activities. Such activities can be represented by BPMN models/patterns.

When a user wishes to use a letter of credit for a purchase of a good, the user searches the repository for the BPMN pattern representing the letter of credit. The activities include shipment that includes transport, insurance, and customs activities supported by appropriate documents. However, transport activity may depend on the geographical locations crossed by the transport, and different geographical locations may have implications on the types of transport allowed, and insurance and customs requirements. Our concept is that the developer is be able to choose appropriate patterns for such activities that depend on geographical locations. If the shipment is a sub-transaction represented by a BPMN pattern, then the developer is able to replace that pattern, at the BPMN level, by an appropriate one that is customized for the geographical location in question, and then incorporate that pattern into the BPMN model for the letter of credit.

### *Smart Contract as a Service (SCaaS)*

By supporting automated creation of smart contracts from BPMN models and providing support for augmentation of BPMN models with BPMN patterns and replacement of patterns in BPMN models with similar patterns, we are striving to create an environment to provide a relatively new concept of Smart-Contract-as-a-Service (SC-as-a-service). In short, the developer would be able to search a repository for BPMN models/patterns for major activities, such as a letter of credit, customize the BPMN model by replacing patterns representing transactions or sub-transactions, with similar patterns for customization purposes to suit the specific context, and then use the TABS+ tool to transform the BPMN model into a smart contract and deploy it on the blockchain selected by the developer.

Thus far there is no authoritative definition of what a SCaaS is. It appears that any support provided for the development of smart contract may be classified as SCaaS. For instance, frameworks used to support development of smart contract, such as (Truffle Suite [53]) for Ethereum and (Hyperledger Composer [24]) for HLF, may be classified as SCaaS. In addition, it appears that any services associated with the smart contract development may be classified CSaaS. For instance, Bitregalo® [2], which is scheduled to provide its SCaaC in 2024, plans to provide a repository of smart contract templates from which a user can choose and then adapt to her/his needs and then deploy on a blockchain for execution. Initially they plan to provide three smart contracts: one, deployed on their blockchain, for authentication using a non-custodial wallet, and two smart contracts to support the tender and escrow, with the former using the services of the latter. At the time of writing, we were unable to ascertain how a smart contract template can be customized or which target blockchain may be used for the smart contract deployment.

Another platform, Simba, provides support for data identity, secure data exchange, and third-party verification. The services are provided by the Simba Chain platform (Simbachain.com [47]) that accepts the smart contract as input together with information on configuration that is facilitated by a smart contract on the Simbachain blockchain. Tools are provided for searching a repository and select services for appropriate smart contract template, while also supporting off-chain storage on Azure platform. Smart contracts deployment targets thus far only Ethereum-type blockchains, such as Quorum and Stellar.

## 5.4 Concluding Remarks

This paper described how the TABS approach, to automated creation of smart contracts from BPMN models, is extended to provide automated support for nested trade transactions that involves several participants. The ultimate objective is to ease the developer's task to create smart contracts for trade transactions. Providing a much simpler process for creation, testing, and deployment of trade smart contracts, we are hoping to decrease the cost of cost and difficulties in creation of smart contracts and spur adoption of smart contracts in general trade and finance of goods and services.

## 6 HISTORY DATES

In case of submissions being prepared for Journals or PACMs, please <u>add history dates after References</u> as (*please note revised date is optional*):

Received June 2023; revised August 2023; …

REFERENCES

[1]    About the Business Process Model and Notation Specification 2.0. (2010). Retr. 2024/02/15 https://www.omg.org/spec/bpmn/2.0/About-

BPMN.

[2]     Belchior, R., Vasconcelos, A., Guerreiro, S., & Correia, M. (2021). A Survey on Blockchain Interoperability: Past, Present, and Future Trends. ACM Computing Surveys, Vol. 4, Issue 8. Article No. 168, pp 1–41. Retr. 2024/02/15 https://doi.org/10.1145/3471140.

[3]     Bitregalo (n.d.). Simplify Contract Management with Bitregalo'sSmart Contract as a Service (SCaaS). Retr. 2023/20/30 https://www.bitregalo.com/scaas.

[4]     Bodorik, P., G. Liu, C., & Jutla, D. (2021). Using FSMs to Find Patterns for Off-Chain Computing. Proc. 3rd Int. Conf. on Blockchain Technology, 2021, 28–34. Retr. 2024/02/15 https://doi.org/10.1145/3460537.3460565.

[5]     Bodorik, P., Liu, C. G., & B Jutla, D. (2023). TABS: Transforming automatically BPMN models into blockchain smart contracts. Elsevier Journal of Blockchain: Research and Applications, 100115, pp. 1-26. Retr. 2024/02/15 https://doi.org/10.1016/j.bcra.2022.100115.

[6]     Bostock, M., Ogievetsky, V., and Heer, J. (2011). Data-Driven Documents. In IEEE Transactions on Visualization and Computer Graphics, Vol 17, Issue 12. pp. 2301-2309.

[7]     BPMN 2.0 Introduction - Flowable Open-Source Documentation. (n.d.). Retrieved Oct. 3, 2023, Retr. Oct 30, 2023, from: https://flowable.com/open-source/docs/.

[8]     BPMN 2.0 Symbols—A complete guide with examples. (n.d.). Camunda. Retr. 2024/02/15  https://camunda.com/bpmn/reference/.

[9]     Business Process Model and Notation (BPMN), Version 2.0.2. (n.d.) Retr. 2024/02/15 https://www.omg.org/spec/BPMN/2.0/PDF.

[10]    Buterin, V. (2015). Ethereum White Paper: A Next Generation Smart Contract & Decentralized Application Platform. Retr. 2024/02/15 https://blockchainlab.com/pdf/Ethereum_white_paper-a_next_generation_smart_contract_and_decentralized_application_platform-vitalik-buterin.pdf.

[11]    Camunda (n.d.). Process Orchestration for end-to-end automation. Retr. 2024/02/15 https://camunda.com.

[12]    Cassandras, C. (1993). Discrete event systems: Modeling and performance analysis. CRC Press. 1st ed. ISBN 10: 0256112126

[13]    Chaum, D. (1983). Blind Signatures for Untraceable Payments. In D. Chaum, R. L. Rivest, & A. T. Sherman (Eds.), Advances in Cryptology. Springer US. pp. 199–203. Retr. 2024/02/15 https://doi.org/10.1007/978-1-4757-0602-4_18.

[14]    Deehan, N. (2021). How Developing Java Apps is Easier with a Process Engine. Camunda. Retr. 2024/02/15 https://camunda.com/blog/2021/11/how-developing-java-apps-is-easier-with-a-process-engine/.

[15]    Di Ciccio C., Cecconi A., Dumas M., García-Bañuelos L., López-Pintado O., Lu Q., Mendling J., Ponomarev A., Tran A. and Weber I. (2019). Blockchain Support for Collaborative Business Processes. Informatik Spektrum Journal, Vol. 42, pages 182–190.

[16]    Dijkman, R., Dumas, M., & Ouyang, C. (2008). Semantics and Analysis of Business Process Models in BPMN. Elsevier Journal on Information and Software Technology, Vol. 50, Issue 12, pp. 1281–1294. Retr. 2024/02/15 https://doi.org/10.1016/j.infsof.2008.02.006.

[17]    Ethereum Average Block Size. (n.d.). Retr. 2024/02/15 https://ycharts.com/indicators/ethereum_average_block_size.

[18]    Ethereum Virtual Machine (EVM). (2023). Retr. 2024/02/15 https://ethereum.org/developers/docs/evm.

[19]    Garcia-Garcia, J. A., Sánchez-Gómez, N., Lizcano, D., Escalona, M. J., & Wojdyński, T. (2020). Using Blockchain to Improve Collaborative Business Process Management: Systematic Literature Review. IEEE Access, Vol. 8, pp. 142312–142336. Retr. 2024/02/15 https://doi.org/10.1109/ACCESS.2020.3013911.

[20]    Girault, A., Lee, B., & Lee, E. A. (1999). Hierarchical finite state machines with multiple concurrency models. IEEE Transactions on Computer-Aided Design of Integrated Circuits and Systems, 18(6), 742–760. Retr. 2024/02/15 https://doi.org/10.1109/43.766725.

[21]    Graphviz (n.d.). Graph Visualization Software Documentation. Retr. 2024/02/15 https://graphviz.org/documentation/.

[22]    Harel, D. (1987). Statecharts: A visual formalism for complex systems. Elsevier Journal of Science of Computer Programming, Vol. 8, Issue 3, pp. 231–274. Retr. 2024/02/15 https://doi.org/10.1016/0167-6423(87)90035-9.

[23]    Hoare, C. A. R. (1978). Communicating sequential processes. Communications of the ACM, Vol. 21, Issue 8, pp. 666–677. Retr. 2024/02/15 https://doi.org/10.1145/359576.359585.

[24]    Hyperledger. (n.d.) Hyperledger Composer. Retr. 2024/02/15 https://hyperledger.github.io/composer/latest/.

[25]    Ismael. (2019). Gas cost of a sha256 hash. Ethereum Stack Exchange. Retr. 2024/02/15 https://ethereum.stackexchange.com/a/76114

[26]    Khan, S. N., Loukil, F., Ghedira-Guegan, C., Benkhelifa, E., & Bani-Hani, A. (2021). Blockchain smart contracts: Applications, challenges, and future trends. Peer-to-Peer Networking and Applications, 14(5), pp. 2901–2925. Retr. 2024/02/16 https://doi.org/10.1007/s12083-021-01127-0.

[27]    Lauster, C., Klinger, P., Schwab, N., & Bodendorf, F. (2020). Literature Review Linking Blockchain and Business Process Management. Smart Cities 6(3), 2023, pp. 1254-1278. Retr. 2024/02/16 https://www.mdpi.com/2624-6511/6/3/61.

[28]    Levasseur, O., Iqbal, M., & Matulevičius, R. (2021). Survey of Model-Driven Engineering Techniques for Blockchain-Based Applications. Proc. 14th IFIP WG 8.1 Working Conference on the Practice of Enterprise Modelling, 2021. Retr. 2024/02/16 https://ceur-ws.org/Vol-3045/paper02.pdf.

[29]    Liu, C., Bodorik, P., & Jutla, D. (2021a). A Tool for Moving Blockchain Computations Off-Chain. Proc. 3rd ACM Int. Symp. on Blockchain and Secure Critical Infrastructure, 103–109. Retr. 2024/02/16 https://doi.org/10.1145/3457337.3457848

[30]    Liu, C., Bodorik, P., & Jutla, D. (2021b). BPMN to smart contracts on blockchains: Transforming BPMN to DE-HSM multi-modal model. Proc. 2021 Int. Conf. on Engineering and Emerging Technologies (ICEET), 1–7. Retr. 2024/02/16 https://doi.org/10.1109/ICEET53442.2021.9659771.

[31]    Liu, C. G., Bodorik, P., & Jutla, D. (2021c). Automating Smart Contract Generation on Blockchains Using Multi-modal Modeling. Journal of Advances on information Technology (JAIT), Vol. 13(3), pp. 213–223. Retr. 2024/02/16 https://doi.org/10.12720/jait.13.3.213-223.

[32] Liu, C. G., Bodorik, P., & Jutla, D. (2022). Supporting Long-term Transactions in Smart Contracts. Proc. Int. Conf. on Blockchain Computing and Applications (BCCA), pp. 11–19. Retr. 2024/02/16 https://doi.org/10.1109/BCCA55292.2022.9922193

[33] Liu, C., Bodorik, P., & Jutla, D. (2023). Long-Term Blockchain Transactions Spanning Multiplicity of Smart Contract Methods. Proc. int. Conf. on Blockchain and Trustworthy Systems (BlockSys'2023), Springer, pp. 141–155. Retr. 2024/02/16 https://doi.org/10.1007/978-981-99-8104-5.

[34] López-Pintado, O., Dumas, M., García-Bañuelos, L., & Weber, I. (2019). Giorgini, P., Weber, B. (eds) Advanced Information Systems Engineering. Proc. CAiSE 2019. Lecture Notes in Computer Science(), vol 11483. Springer, Cham. Retr. 2024/02/a6 https://doi.org/10.1007/978-3-030-21290-2_25.

[35] López-Pintado, O., García-Bañuelos, L., Dumas, M., Weber, I., & Ponomarev, A. (2019). CATERPILLAR: A Business Process Execution Engine on the Ethereum Blockchain. Software: Practice and Experience, Vol. 49, Issue 7, pp. 1162-1193. Retr. 2024/02/16 https://doi.org/10.48550/arXiv.1808.03517.

[36] Loukil, F., Boukadi, K., Abed, M., & Ghedira-Guegan, C. (2021). Decentralized collaborative business process execution using blockchain. World Wide Web, 24(5), 1645–1663.

[37] Marr, B. (2017). A Short History of Bitcoin and Crypto Currency Everyone Should Read. Forbes Newsletter. Retr. 2024/02/16 https://www.forbes.com/sites/bernardmarr/2017/12/06/a-short-history-of-bitcoin-and-crypto-currency-everyone-should-read/.

[38] Mavridou, A., & Laszka, A. (2018a). Designing Secure Ethereum Smart Contracts: A Finite State Machine Based Approach. Proc. Int. Conf. Financial Cryptography and Data Security. Springer LNCS, Vol. 10957, pp. 523-540. Retr. 2024/02/16 https://doi.org/10.48550/arXiv.1711.09327.

[39] Mavridou, A., & Laszka, A. (2018b). Tool Demonstration: FSolidM for Designing Secure Ethereum Smart Contracts. Proc. Principles of Security and Trust (POST 2018). Springer LNCS, Vol. 10804, pp. 217-277. Retr. https://link.springer.com/book/10.1007/978-3-319-89722-6 https://link.springer.com/book/10.1007/978-3-319-89722-6.

[40] Mendling, J., Weber, I., Aalst, W. V. D., Brocke, J. V., Cabanillas, C., Daniel, F., Debois, S., Ciccio, C. D., Dumas, M., Dustdar, S., Gal, A., García-Bañuelos, L., Governatori, G., Hull, R., Rosa, M. L., Leopold, H., Leymann, F., Recker, J., Reichert, M., … Zhu, L. (2018). Blockchains for Business Process Management—Challenges and Opportunities. ACM Transactions on Management Information Systems, 9(1), 1–16. Retr. 2024/02/16 https://doi.org/10.1145/3183367.

[41] Nabben, K. (2022). Decentralized Technology in Practice: Social and technical resilience. Proc. 42nd Int. Conf. on Distributed Computing Systems Workshops (ICDCSW), pp. 66–72. Retr. 2024/10/16 https://ieeexplore.ieee.org/document/9951370.

[42] Nakamoto, S. (2008). Bitcoin: A Peer-to-Peer Electronic Cash System. Originally published as a white paper on a mailing list metzdowd.com on Oct. 31, 2008. Retr. 2024/10/16 https://bitcoin.org/bitcoin.pdf.

[43] Node.js – Open-source, cross-platform JavaScript runtime environment. (n.d.). Retr. 2024/10/16 https://nodejs.org/en.

[44] InterPlanetary File System (IPFS): Concepts. (n.d.). Retr. 2024/02/16 https://docs.ipfs.tech/concepts/.

[45] Ethereum's New 1MB Blocksize Limit. (2021). In Bitmex Research Blog. Retr. 2024/10/16 https://blog.bitmex.com/ethereums-new-1mb-blocksize-limit/.

[46] Saito, K., & Yamada, H. (2016). What's So Different about Blockchain? — Blockchain is a Probabilistic State Machine. Proc. 36th Int. Conf. on Distributed Computing Systems Workshops (ICDCSW), pp. 168–175. Retr. 2024/10/16 https://doi.org/10.1109/ICDCSW.2016.28.

[47] Simbachian.com (n.d.). Simba: Smart-Contract-as-a-Service User Guide. Retr. 2024/10/16 https://simbachain.com/wp-content/uploads/2019/08/SIMBA-SCaaS-User-Guide-v1.0.2.pdf.

[48] Steichen, M., Fiz Pontiveros, B., Norvill, R., Shbair, W., & State, R. (2018). Blockchain-Based, Decentralized Access Control for IPFS. Proc. 2018 IEEE Int. Conf. on iThings and GreenCom and CPSCom and SmartData, pp. 1499-1506. Retr. 2024/02/16 https://doi.org/10.1109/Cybermatics_2018.2018.00253.

[49] Szabo, N. (1996). Smart Contracts: Building Blocks for Digital Free Markets. Extropy Journal of Transhuman Thought, 16.

[50] Taylor, P. J., Dargahi, T., Dehghantanha, A., Parizi, R. M., & Choo, K.-K. R. (2020). A systematic literature review of blockchain cyber security. Digital Communications and Networks, 6(2), 147–156. Retr. 2024/10/16 http://doi.org/10.1016/j.dcan.2019.01.005.

[51] Tran, A., Lu, Q., & Weber I. (2018). Lorikeet: A Model-Driven Engineering Tool for Blockchain-Based Business Process Execution and Asset Management. Proc. 2018 Int. Conf. on Business Process Management, pp. 1-5. Retr. 2024/02/16 https://api.semanticscholar.org/CorpusID:52195200.

[52] Dikmans, L. (2008). Transforming BPMN into BPEL: Why and How. Oracle Middleware/Technical Details/Technical Article. Retr. 2024/10/16 https://www.oracle.com/technical-resources/articles/dikmans-bpm.html.

[53] Truffle: The most comprehensive suite of tools for smart contract development. (n.d.). Truffle Suite Documentation. Retr. 2024/10/16 https://trufflesuite.com.

[54] Vacca, A., Di Sorbo, A., Visaggio, C. A., & Canfora, G. (2021). A systematic literature review of blockchain and smart contract development: Techniques, tools, and open challenges. Journal of Systems and Software, Vol. 174, 110891. Retr. 2024/10/16 https://doi.org/10.1016/j.jss.2020.110891.

[55] Weber, I., Xu, X., Riveret, R., Governatori, G., Ponomarev, A., & Mendling, J. (2016). Untrusted Business Process Monitoring and Execution Using Blockchain. Proc. Int. Conf. 2016 on Business Process Management, pp. 329–347, M. La Rosa, P. Loos, & O. Pastor (Eds.), Springer International Publishing. Retr. 2024/10/16 https://doi.org/10.1007/978-3-319-45348-4_19.

[56] Yannakakis, M. (2000). Hierarchical State Machines. Proc. IFIP Int. Conf. on Theoretical Computer Science, Exploring New Frontiers of

Theoretical Informatics, pp. 315–330, ACM.